%% file: Main.tex
\documentclass[pdflatex,sn-mathphys-num]{sn-jnl}% Math and Physical Sciences Numbered Reference Style
\usepackage[normalem]{ulem}
\usepackage{graphicx}%
\usepackage{multirow}%
\usepackage{amsmath,amssymb,amsfonts}%
\usepackage{amsthm}%
\usepackage{mathrsfs}%
\usepackage[title]{appendix}%
\usepackage{xcolor}%
\usepackage{textcomp}%
\usepackage{manyfoot}%
\usepackage{booktabs}%
\usepackage{algorithm}%
\usepackage{algorithmicx}%
\usepackage{algpseudocode}%
\usepackage{listings}%
\usepackage{pdfpages}

\theoremstyle{thmstyleone}%
\theoremstyle{thmstyletwo}%

\theoremstyle{thmstylethree}%

\begin{document}

\title[Article Title]{Spectral dynamics reservoir computing for high-speed hardware-efficient neuromorphic processing}

%%=============================================================%%
%% GivenName	-> \fnm{Joergen W.}
%% Particle	-> \spfx{van der} -> surname prefix
%% FamilyName	-> \sur{Ploeg}
%% Suffix	-> \sfx{IV}
%% \author*[1,2]{\fnm{Joergen W.} \spfx{van der} \sur{Ploeg} 
%%  \sfx{IV}}\email{iauthor@gmail.com}
%%=============================================================%%

\author*[1]{\fnm{Jiaxuan} \sur{Chen}}\email{CHEN.Jiaxuan@nims.go.jp}

\author*[1]{\fnm{Ryo} \sur{Iguchi}}\email{IGUCHI.Ryo@nims.go.jp}

\author[1,2]{\fnm{Sota} \sur{Hikasa}}\email{HIKASA.Sota@nims.go.jp}

\author[1,2]{\fnm{Takashi} \sur{Tsuchiya}}\email{TSUCHIYA.Takashi@nims.go.jp}

\affil*[1]{\orgdiv{Research Center for Materials Nanoarchitectonics (MANA)}, 
\orgname{National Institute for Materials Science (NIMS)}, 
\orgaddress{\street{1-1 Namiki}, \city{Tsukuba}, \state{Ibaraki}, \postcode{305-0044}, \country{Japan}}}

\affil[2]{\orgdiv{Faculty of Advanced Engineering}, \orgname{Tokyo University of Science}, \orgaddress{\street{6-3-1 Niijuku}, \city{Katsushika}, \state{Tokyo}, \postcode{125-8585}, \country{Japan}}}

%%==================================%%
%% Sample for unstructured abstract %%
%%==================================%%

\abstract{Physical reservoir computing~(PRC) is a promising brain-inspired computing architecture for overcoming the von Neumann bottleneck by utilizing the intrinsic dynamics of physical systems. However, a major obstacle to its real-world implementation lies in the tension between extracting sufficient information for high computational performance and maintaining a hardware-feasible, high-speed architecture. Here, we report spectral dynamics reservoir computing (SDRC), a broadly applicable framework based on analogue filtering and envelope detection that bridges this gap. SDRC effectively exploits the fast spectral dynamics embedded in short-time, coarse spectra of material responses to attain strong computational capability while maintaining high-speed processing and minimal hardware overhead. This approach circumvents the need for implementation-intensive, precision-sensitive integrated circuits required in high-speed time-multiplexing measurements, while enabling real-time use of the material's spectral manifold as a high-dimensional computational resource. We implement and experimentally demonstrate SDRC applied to spin waves that achieves state-of-the-art-level performance with only 56 nodes on benchmark tasks of parity-check and second-order nonlinear autoregressive moving average, as well as high accuracy of 98.0\% on a real-world problem of speech recognition.}

% \keywords{keyword1, Keyword2, Keyword3, Keyword4}

%%\pacs[JEL Classification]{D8, H51}

%%\pacs[MSC Classification]{35A01, 65L10, 65L12, 65L20, 65L70}

\maketitle

\section{Introduction}\label{sec:intro}

\begin{figure}
\centering
\includegraphics[width=1\textwidth]{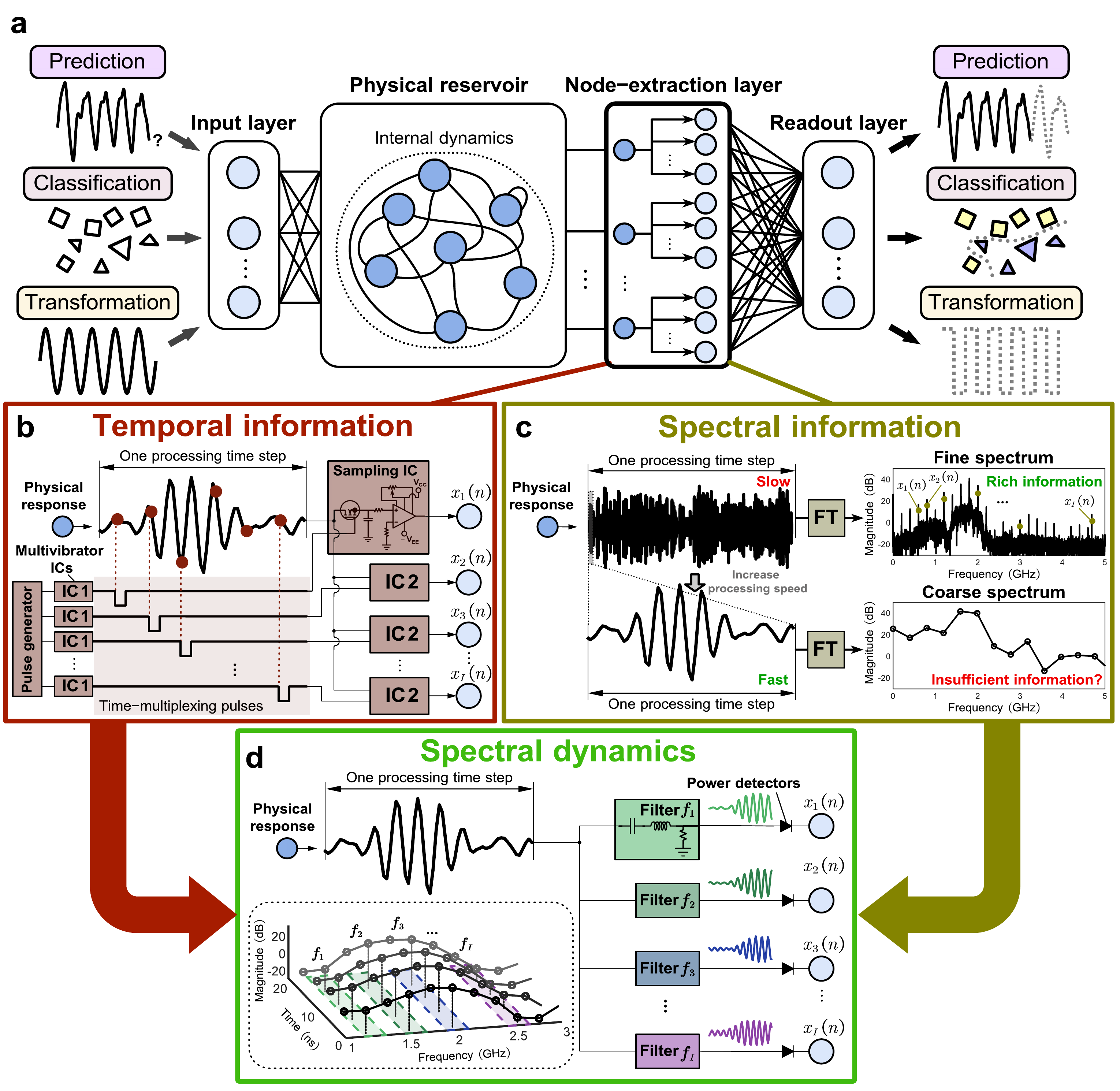}
\caption{\textbf{\textbar\ PRC architecture and different information-extraction schemes.}
\textbf{a}, Typical architecture of a PRC system capable of performing various information processing tasks. A node-extraction layer is employed to extract high-dimensional information for effective computation. \textbf{b}, A typical scheme of temporal information extraction, where a pulse generator first produces narrow time pulses, which are distributed to multiple multivibrator ICs to introduce high-precision temporal delays, thereby forming time-multiplexing sampling references. Multiple sampling ICs are then used to sample and hold the temporal evolution of a physical response according to the time-multiplexing pulses as clock signals. \textbf{c}, A typical scheme of spectral information extraction, where a finely resolved frequency spectrum is obtained through the Fourier transform~(FT), necessitating a long integration time window that inherently introduces computational overhead and operation latency that hinder true real-time processing. Increasing processing speed results in a coarse spectrum where distinctive peaks are blurred. Whether such a spectrum retains sufficiently rich latent information for complex neuromorphic computation remains a question. \textbf{d}, A scheme of spectral dynamics extraction, where multiple analogue filters extract short-time, coarse spectral dynamics of the physical response with power detectors for envelope detection, providing a high-speed and hardware-efficient implementation for real-time computation.}
\label{fig:SDRC_flow}
\end{figure}

Fundamental bandwidth and energy limitations of conventional von Neumann architectures have motivated intensive research into neuromorphic computing for brain-inspired, highly parallel, and power-efficient information processing~\cite{wright2022deep,shastri2021photonics,grollier2020neuromorphic,roy2019towards}. Physical reservoir computing~(PRC), as a prominent example of neuromorphic architectures, utilizes nonlinear dynamics of physical systems to achieve efficient and complex information processing (Fig.~\ref{fig:SDRC_flow}a)~\cite{yan2024emerging,tanaka2019recent,appeltant2011information,jaeger2001echo}, meeting the demands of modern real-time and data-intensive applications~\cite{markovic2020physics,yan2024emerging,zidan2018future}. As a result, PRC has been actively explored across a wide range of material platforms, including photonic~\cite{aadhi2025scalable,rafayelyan2020large,sunada2019photonic,larger2017high,brunner2013parallel}, electronic~\cite{nishioka2025two,zhong2022memristor,liu2022multilayer,nishioka2022edge,zhong2021dynamic}, and spintronic systems~\cite{gartside2022reconfigurable,lee2024task,prychynenko2018magnetic,chen2025spintronic,tsunegi2019physical}. 

Despite these advances, there is still a critical gap from conceptual validation to real-world implementation of PRC, stemming from the challenge of achieving high-performance information processing while maintaining a hardware-feasible and real-time system architecture~\cite{markovic2020physics,yan2024emerging,tanaka2019recent}. This challenge is rooted in the fact that extracting sufficiently high-dimensional information from the materials during rapid processing typically necessitates complex and implementation-heavy extraction schemes (Fig.~\ref{fig:SDRC_flow}a). Consequently, there is an urgent need for more hardware-friendly approaches that enable effective information extraction~\cite{kurebayashi2026metrics}.

Overcoming these hardware barriers requires a deeper exploitation of the intrinsic dynamics that define the physical substrates. Many physical materials inherently exhibit nonlinear oscillatory dynamics that can encode dense computational information~\cite{landau1980statistical,strogatz2024nonlinear}. Because these oscillatory features are intertwined within their temporal responses, resolving them purely through time-domain measurements yields limited information and imposes substantial hardware demands, especially at elevated operating speeds (Fig.~\ref{fig:SDRC_flow}b). Therefore, information from materials' well-resolved frequency spectra has been considered as an alternative computational resource~\cite{korber2023pattern,lee2024task,butschek2022photonic,lupo2023deep}. However, high-speed utilization of spectral information is fundamentally discouraged by the time-frequency trade-off (Fig.~\ref{fig:SDRC_flow}c), where increasing processing speeds inevitably leads to a coarse, discretized frequency spectrum. Hence, the real-time feasibility of spectral-domain-based computing hinges on the open question of whether fast spectral dynamics from unresolved spectra can still support complex neuromorphic computation.

In this Article, we report that a spectral dynamics reservoir computing~(SDRC) system composed of analogue filtering and envelope detection (Fig.~\ref{fig:SDRC_flow}d) shows strong computational capability with only a small number of readout channels (nodes). SDRC extracts dense spectral dynamics from coarse spectra with simple circuitry and basic electronic components, thereby circumventing the hardware overhead of high-precision, tightly synchronized clocking and complex sampling integrated circuits (ICs) required by traditional time-multiplexed architectures (Fig.~\ref{fig:SDRC_flow}b), as commonly used in electronic devices~\cite{appeltant2011information,namiki2025iono,watt2020reservoir,larger2017high,zhong2021dynamic,brunner2013parallel}. As a platform-agnostic framework, we demonstrate its hardware implementation applied to spin waves---oscillatory dynamics of magnetic order with inherent nonlinearity---and reveal a physics-based correspondence between high computational performance and the spectral bands of spin waves. Guided by this insight for information-effective node optimization, we experimentally realize an SDRC system with only $56$ nodes that achieves state-of-the-art-level performance on parity-check and second-order nonlinear autoregressive moving average (NARMA-$2$) benchmarks, as well as high accuracy in a real-world speech-recognition problem. 

\section{Spectral dynamics reservoir computing using spin waves}\label{sec:SDRC}

Spin waves are collective precessional motions of electron spins in magnetically ordered materials. They naturally possess rich spectral dynamics arising from their dispersive characteristics and nonlinear modal interactions within a single magnetic medium~\cite{gurevich2020magnetization,bauer2015nonlinear,krivosik2010hamiltonian,stancil2009spin,chen2026nonlinear}, making them attractive candidates for compact, energy-efficient, and high-speed edge computing devices~\cite{kurebayashi2026metrics,chumak2022advances,mahmoud2020introduction,nakane2018reservoir,namiki2023experimental,chen2025analytic}. In the linear-regime approximation, spin waves at different frequencies propagate independently and can be treated as distinct dynamical modes with different group velocities~\cite{gurevich2020magnetization,stancil2009spin}. Beyond the linear regime, the nonlinear nature of spin waves gives rise to inter-modal interactions. Spin waves can be approximated as a superposition of complex exponentials $\sum_{n} \mathbf{M}_n e^{i\left( \mathbf{k}_n\cdot \mathbf{r}- \omega_n t\right)}$, with $\mathbf{M}_n$, $\mathbf{k}_n$, and $\omega_n$ the amplitude vector, wavevector, and angular frequency of the $n$-th spin-wave mode, $\mathbf{r}$ the position vector, and $i$ the imaginary unit. Then, the coupled dynamics of these modes satisfy
\begin{equation}
\sum_{n} i\omega_n\mathbf{M}_n e^{i\left( \mathbf{k}_n\cdot \mathbf{r}- \omega_n t\right)}
= \gamma \, \left(\sum_{n} \mathbf{M}_n e^{i\left( \mathbf{k}_n\cdot \mathbf{r}- \omega_n t\right)}\right) \times \left(\sum_{n} \hat{\mathcal G}(\mathbf{k}_n)\mathbf{M}_n e^{i\left( \mathbf{k}_n\cdot \mathbf{r}- \omega_n t\right)} + \mathbf{B}_0\right),
\label{eq:LLG_partial}
\end{equation}
where $\gamma$ is the gyromagnetic ratio, $\hat{\mathcal G}(\mathbf{k})$ the combined exchange-dipolar Green's function in reciprocal space, and $\mathbf{B}_0$ is the bias field vector (Supplementary Note 1). As manifested in this relation, nonlinear interactions between two modes at frequencies $\omega_{n_1}$ and $\omega_{n_2}$ result in additional spectral components such as $\omega_{n_1}\pm\omega_{n_2}$, enriching the densely intertwined nonlinear dynamics in the spin-wave spectrum.

\begin{figure}
\centering
\includegraphics[width=1\textwidth]{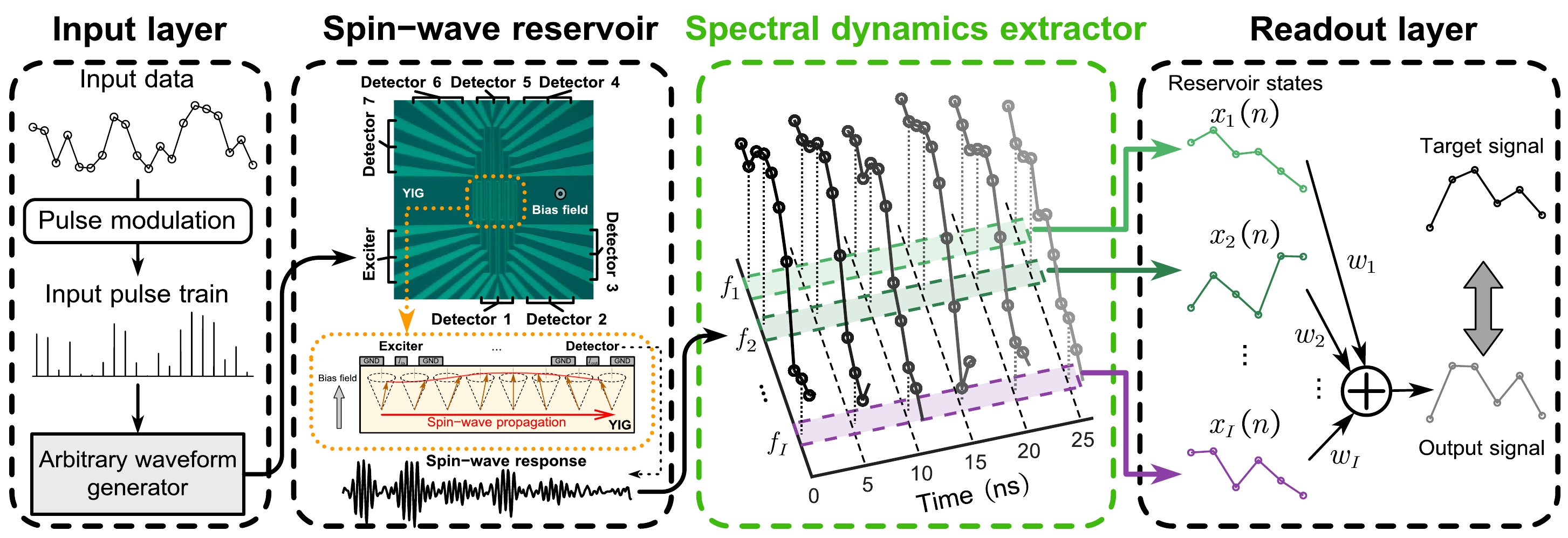}
\caption{\textbf{\textbar\ The SDRC system using spin-wave dynamics.} An analogue pulse train that encodes the input data through pulse modulation is fed into a spin-wave reservoir made by a YIG single crystal to excite spin-wave propagation. The resulting spin-wave responses are subsequently processed to extract parallel spectral dynamics as reservoir states from their coarse spectra through analogue filtering and envelope detection. These high-dimensional reservoir states are then combined through a linear readout with trainable weights to generate output signals for information processing tasks.}\label{fig:SWRC}
\end{figure}

To exploit this dynamical richness of spin waves in computation, we construct the SDRC system as illustrated in Fig.~\ref{fig:SWRC}. An yttrium iron garnet (YIG) single crystal with $(111)$-orientation is used as the spin-wave reservoir, patterned with ten coplanar waveguides (CPWs), among which the second CPW from the left is used as the spin-wave exciter and seven as detectors (Methods). An out-of-plane bias magnetic field is applied to the YIG medium to establish forward volume spin waves (FVSWs). An analogue pulse train that encodes input data is fed into the exciter for broadband spin-wave excitation within the YIG medium. We probe the latent dense spectral manifold of spin-wave responses measured at each detector through analogue filtering and envelope detection at various spectral bands, extracting spectrally masked temporal dynamics as reservoir states, termed spectral nodes (Fig.~\ref{fig:SWRC}). These extracted reservoir states are then fed into a linear readout and combined with a set of trainable weights to generate output signals and perform computational tasks. 

\section{Benchmark performance of SDRC with spectral-node emulation}\label{sec:emulation_benchmark}

To characterize the information distribution in the spectral dynamics, we first examine the SDRC system by measuring spin-wave responses with an oscilloscope and processing them via software emulation of band-pass filtering and envelope detection (Methods). We use spin-wave responses obtained at a bias magnetic field of $187.3$~mT, where dominant spin-wave modes are established with frequencies around $2$~GHz (Supplementary Fig.~1). For each detector, $50$ spectral nodes are emulated with their center frequencies ranging from $0.1$ to $5$~GHz in increments of $0.1$~GHz to fully cover the spin-wave spectrum. Each discrete symbol is processed at a rate of $0.2$~GHz, which approaches the operating spin-wave frequency. We evaluate the system performance using parity check~\cite{tanaka2022self,lee2023physical,toprasertpong2022reservoir,rajib2025magneto,heins2025benchmarking,nagase2024spin,tsunegi2019physical} and NARMA-$2$~\cite{nishioka2025two,namiki2024opto,akashi2020input,kan2021simple,akai2022performance,yamada2024physical,kan2022physical}, as two widely adopted benchmark tasks in the studies of PRC. To illustrate the information effectiveness of SDRC, we compare its performance with the conventional time-multiplexing (virtual-node) approach~\cite{appeltant2011information} using the same experimentally measured spin-wave responses (Methods). 

\begin{figure}
\centering
\includegraphics[width=1\textwidth]{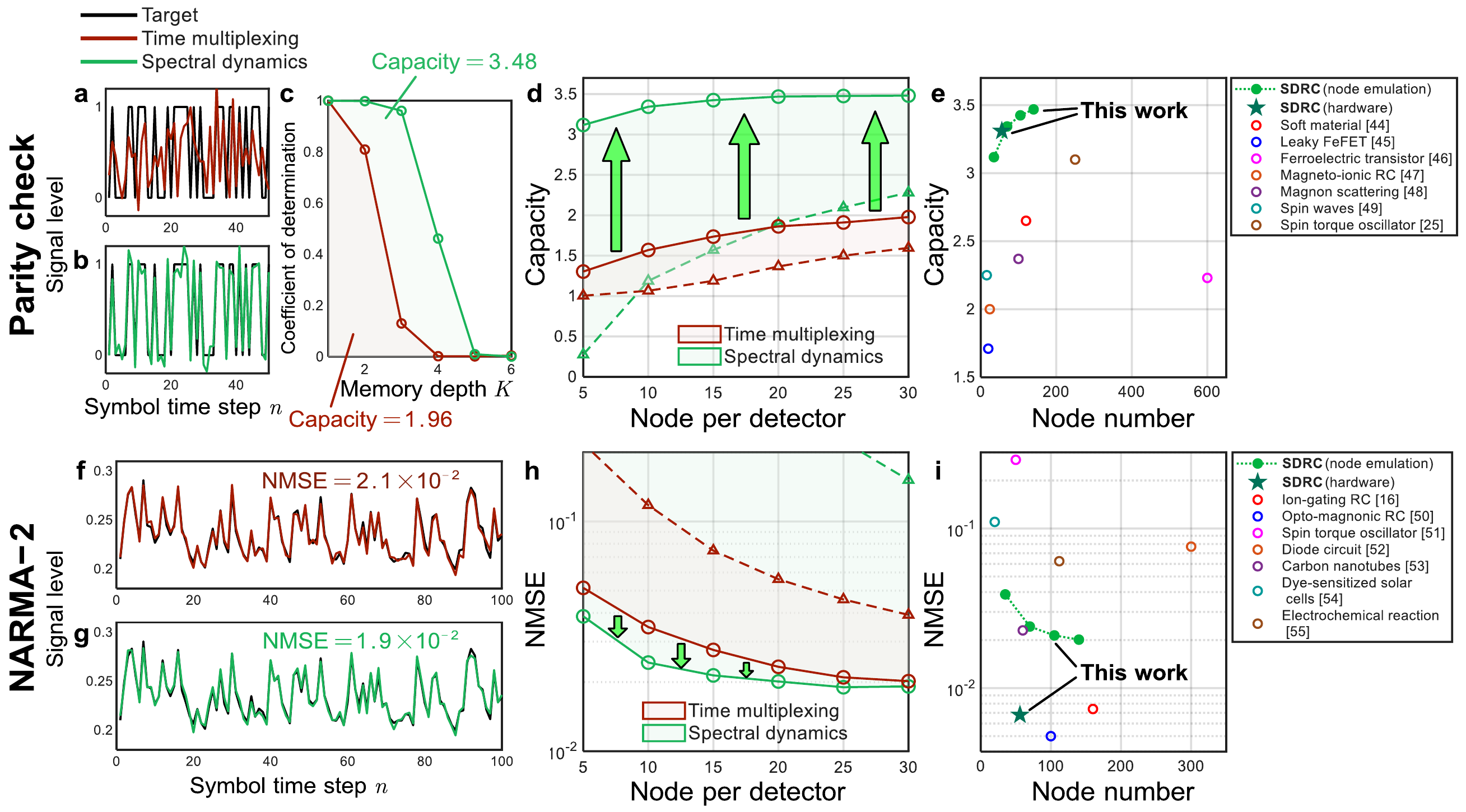}
\caption{\textbf{\textbar\ Benchmark performance of the SDRC system on parity check and NARMA-$2$.} \textbf{a}, \textbf{b}, Example time-domain waveforms for the parity-check task at a memory depth $K = 3$, showing the target signal (black) and the predicted signals obtained using time multiplexing (brown) and spectral dynamics extraction (green), respectively, with $20$ nodes per detector. \textbf{c}, Corresponding coefficients of determination plotted against memory depth $K$, where the enclosed area indicates the parity-check capacity for each approach. \textbf{d}, Parity-check capacity in relation to the number of nodes per detector for both approaches, showing their maximum (circles) and minimum (triangles) values obtained from random node selections. \textbf{e}, Comparison of SDRC performance on parity check in terms of capacity and node number with other PRC~\cite{tanaka2022self,lee2023physical,toprasertpong2022reservoir,rajib2025magneto,heins2025benchmarking,nagase2024spin,tsunegi2019physical}. \textbf{f}, \textbf{g}, Example time-domain waveforms for the NARMA-$2$ task, showing the target signal (black) and the predicted signals obtained using time multiplexing (brown) and spectral dynamics extraction (green), respectively, with $20$ nodes per detector. \textbf{h}, NMSE in relation to the number of nodes per detector for both approaches, showing their minimum (circles) and maximum (triangles) values from random node selections. \textbf{e}, Comparison of SDRC performance on NARMA-$2$ in terms of NMSE and node number with other PRC~\cite{nishioka2025two,namiki2024opto,akashi2020input,kan2021simple,akai2022performance,yamada2024physical,kan2022physical}.}\label{fig:benchmark_performance}
\end{figure}

Parity check requires the system to determine whether the number of ones in a given memory depth (denoted as $K$) of a binary input sequence is even or odd, thereby testing its capability to process temporally distributed information through nonlinear logic. This task is highly relevant to real-world applications such as error detection, temporal classification, and sequence-based decision making. A larger $K$ requires longer memory retention and higher-order nonlinear processing ability of the reservoir. Parity-check capacity is calculated to quantify the system performance. Figures~\ref{fig:benchmark_performance}a and~\ref{fig:benchmark_performance}b show examples of parity prediction at a challenging memory depth $K = 3$ for time multiplexing and spectral dynamics extraction, respectively. By extracting the spectrally embedded nonlinear spin-wave dynamics, SDRC achieves more accurate prediction, whereas the time-multiplexing approach exhibits only limited learning of the target trend. This results in a substantial improvement in the parity-check capacity (Fig.~\ref{fig:benchmark_performance}c), such as an increase from $1.96$ for time multiplexing to $3.48$ for spectral dynamics extraction when using $140$ nodes for both approaches.

To examine the influence of node number on computational performance, random node selections from the emulated node pools are conducted. The resulting capacity ranges for both approaches are shown in Fig.~\ref{fig:benchmark_performance}d, where SDRC consistently yields higher maximal capacity across all examined numbers of nodes per detector, demonstrating improved information effectiveness of this approach. Figure~\ref{fig:benchmark_performance}e compares our results with other reported high-performance PRC systems~\cite{tanaka2022self,lee2023physical,toprasertpong2022reservoir,rajib2025magneto,heins2025benchmarking,nagase2024spin,tsunegi2019physical}, where SDRC with emulated spectral nodes achieves superior capacity with substantially fewer nodes (such as capacities of $3.11$ with $35$ nodes and $3.48$ with $140$ nodes).

NARMA-$2$ is another benchmark task that probes both nonlinearity and memory, requiring the system to learn and predict the evolution of a second-order nonlinear dynamical system. Performance in this task is quantified using the normalized mean squared error~(NMSE) between the predicted and target signals. Figures~\ref{fig:benchmark_performance}f and~\ref{fig:benchmark_performance}g show the target dynamics together with the corresponding predictions from the respective approaches for $140$ nodes. While the time-multiplexing approach captures the target dynamics with $\text{NMSE} = 2.1 \times10^{-2}$, spectral dynamics extraction further improves the prediction accuracy to $\text{NMSE} = 1.9 \times10^{-2}$. The influence of node number on performance is further examined in Fig.~\ref{fig:benchmark_performance}h, where SDRC markedly improves the NMSEs, particularly at smaller node numbers, owing to the richer information extracted by the spectral nodes. Figure~\ref{fig:benchmark_performance}i positions this work with other reported PRC~\cite{tanaka2022self,lee2023physical,toprasertpong2022reservoir,rajib2025magneto,heins2025benchmarking,nagase2024spin,tsunegi2019physical} on the NARMA-2 task, where the NMSEs obtained using emulated spectral nodes are comparable with state-of-the-art results (such as NMSEs of $2.7 \times10^{-2}$ with $35$ nodes and $1.9 \times10^{-2}$ with $140$ nodes). 

Interestingly, as observed in both tasks (Figs.~\ref{fig:benchmark_performance}d and~\ref{fig:benchmark_performance}h), spectral nodes manifest larger performance variations among different node selections, suggesting an uneven distribution of useful information across the spin-wave spectrum. In particular, high performance can be achieved in both tasks even with a small number of nodes (e.g., $5$ per detector) when selecting appropriate spectral nodes. This necessitates a deeper investigation into the underlying principles governing these computationally useful nodes, as establishing such guidelines is essential for efficient node optimization.

\section{Spectral principles for efficient node selection}\label{sec:node_selection}

\begin{figure}
\centering
\includegraphics[width=1\textwidth]{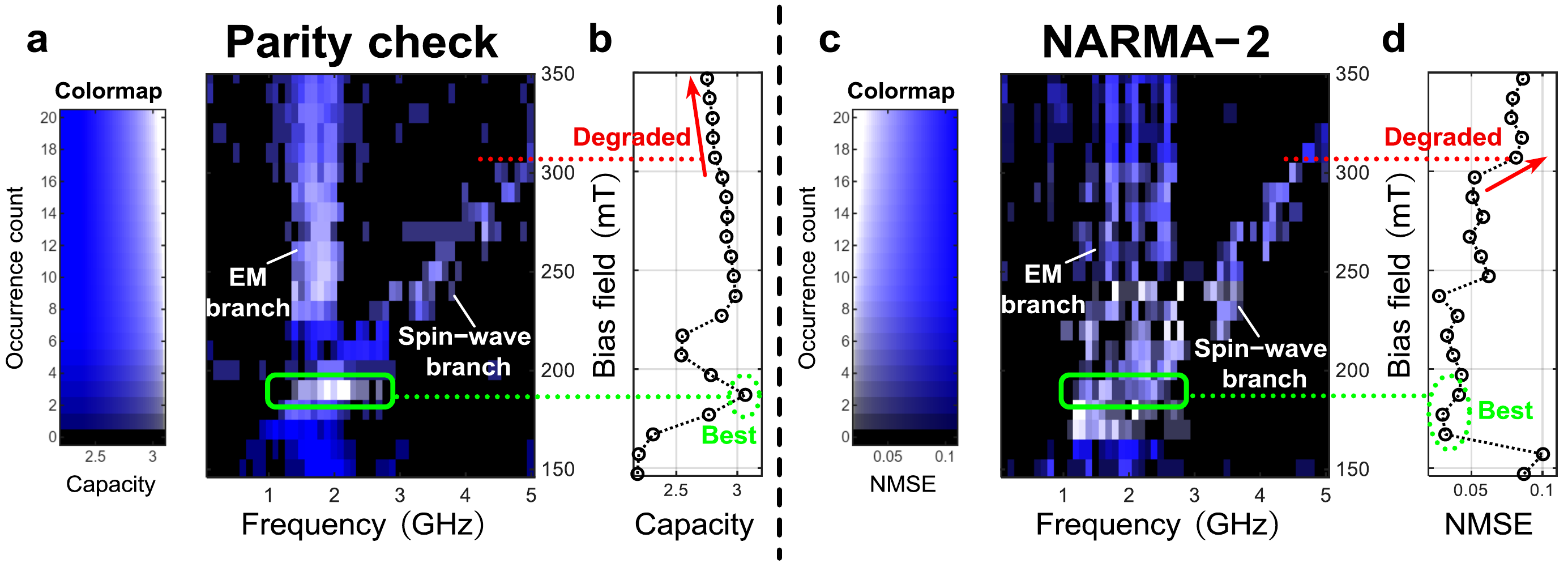}
\caption{\textbf{\textbar\ Spectral distributions of information-effective spectral nodes.} \textbf{a}, \textbf{c}, Occurrence distributions of spectral nodes in relation to the bias magnetic field for the parity-check and NARMA-$2$ tasks, respectively. Occurrence counts are accumulated over the top $20$ performing node combinations obtained from random node selections with $5$ nodes per detector. The colormap jointly encodes the occurrence count by saturation and computational performance (capacity and NMSE, respectively) by brightness. \textbf{b}, \textbf{d}, Bias-field-dependent capacity and NMSE for parity check and NARMA-$2$, respectively, averaged over the top $20$ node selections. }\label{fig:node_dependence}
\end{figure}

To clarify the origin of the information-effective spectral nodes, we analyze how those nodes are influenced with respect to the main frequencies of spin waves, which heavily depend on the strength of the bias magnetic field~\cite{stancil2009spin} (Supplementary Note 1). Specifically, we sweep the field from $147.3$ to $347.5$~mT in steps of approximately $10$~mT, shifting the dominant spin-wave frequency from roughly $1$ to $6$~GHz (Supplementary Fig.~1). For each bias field, we analyze the top $20$ performing spectral-node combinations identified from random node selections in both tasks for the case of $5$ nodes per detector. We count the number of occurrences of individual spectral nodes within the top $20$ combinations, and the resulting node distributions in relation to the bias field are presented in Figs.~\ref{fig:node_dependence}a and~\ref{fig:node_dependence}c for parity check and NARMA-$2$, respectively. The corresponding capacity and NMSE averaged over the $20$ selections are shown in Figs.~\ref{fig:node_dependence}b and~\ref{fig:node_dependence}d. Here, two branches of spectral nodes with high occurrence are consistently observed in both tasks. The first branch consists of spectral nodes with an approximately linear dependence on the bias field, aligning with the field-dependent ferromagnetic resonance (FMR) frequency of spin waves in the device (Supplementary Fig.~1). Spin-wave modes near the FMR frequency naturally exhibit dominant broadband excitation together with dense nonlinear coupling, thereby concentrating more computationally useful information. The other branch comprises field-independent nodes around $2$~GHz, which are attributed to direct electromagnetic~(EM) coupling between the exciter and detector CPWs. The fast and predominantly linear nature of EM waves allows these nodes to provide instantaneous memory of the input information for computation. With a limited node number, optimal performance for both tasks is achieved when the spin-wave branch overlaps with the EM-wave branch at a bias field of around $190$~mT, where both propagating nonlinear features and instant linear memory can be simultaneously extracted within the same spectral nodes. These results strongly demonstrate that spectral dynamics extracted from only a partial band of the coarse spectral manifold can provide sufficient high-dimensional information necessary for complex computation, owing to the densely coupled nonlinearities of spin-wave dynamics.

As the bias field increases beyond $300$~mT, where spin waves start to fall outside the spectral range, we observe a gradual degradation in the capacity of the parity-check task (Fig.~\ref{fig:node_dependence}b) and a rapid increase in the NMSE of the NARMA-$2$ task (Fig.~\ref{fig:node_dependence}d). These trends underscore the contributions of spectral dynamics from spin waves in enabling high computational performance. Furthermore, we observe moderate performance even at fields beyond $300$~mT, which can be attributed to both the intrinsic nonlinearity in the envelope detection process and frequency mixing of spin-wave modes that produce low-frequency spectral components~\cite{gurevich2020magnetization,chen2026nonlinear} (as suggested in equation~\eqref{eq:LLG_partial}). We also observe poor performance at bias fields below $160$~mT, likely because the field strength is insufficient to establish stable spin waves while EM waves persist.

\section{Hardware realization of SDRC}\label{sec:hardware_demo}

\begin{figure}
\centering
\includegraphics[width=1\textwidth]{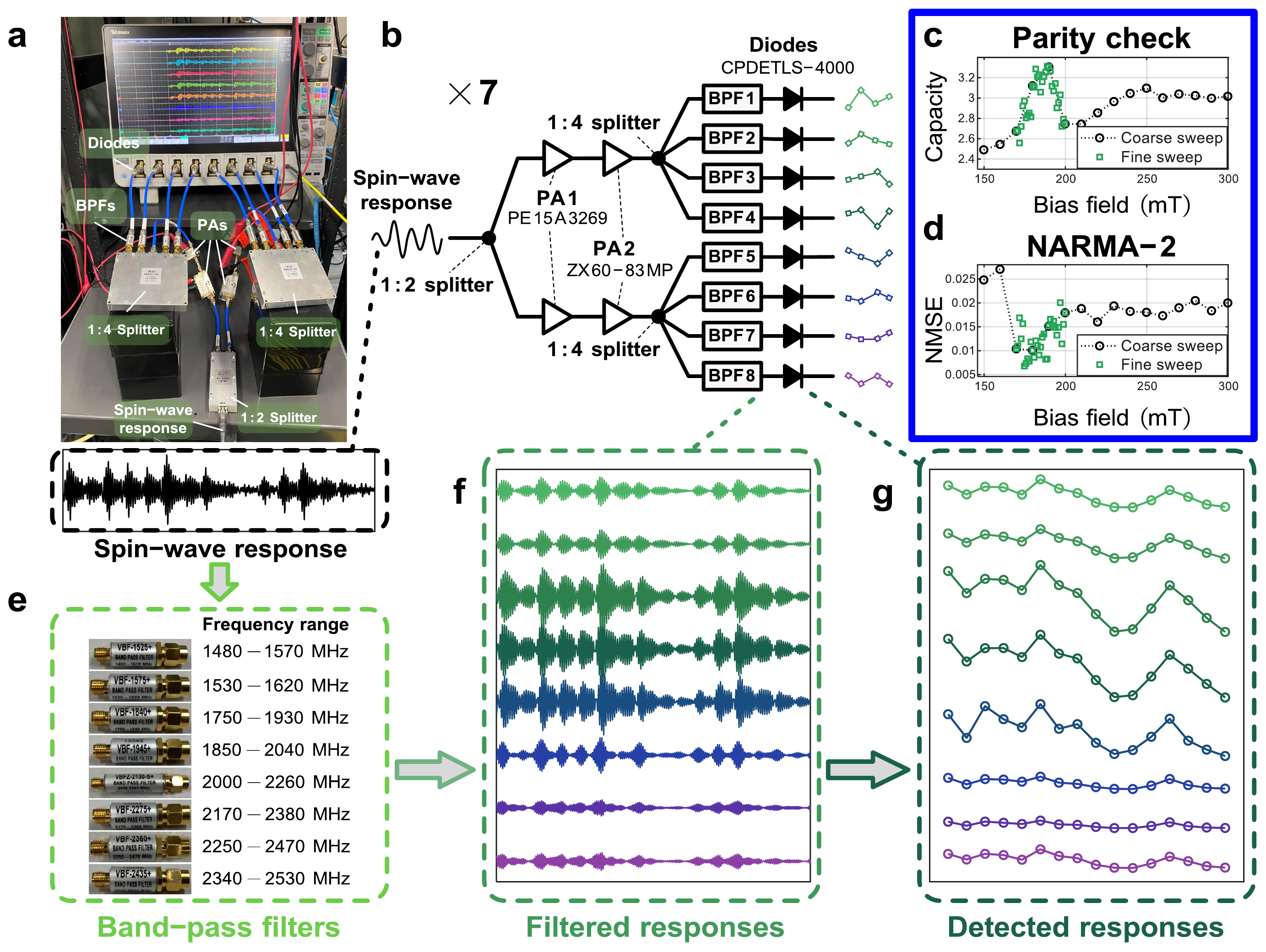}
\caption{\textbf{\textbar\ Hardware demonstration of SDRC.} \textbf{a}, Experimental setup for spectral dynamics extraction, where the spin-wave response from a detector is split into eight channels, each filtered by a BPF with a distinct spectral band. The filtered responses are then rectified and detected using diodes, with waveforms recorded in an oscilloscope. \textbf{b}, Circuit diagram of the spectral dynamics extraction setup for each detector. \textbf{c}, \textbf{d}, Computational performance of the SDRC system with hardware-implemented spectral nodes for the parity-check and NARMA-$2$ tasks, respectively, showing the capacity and NMSE in relation to the bias field. Both a coarse sweep (black) from $149.8$~mT to $299.8$~mT in steps of $10$~mT and a fine sweep (green) from $169.8$~mT to $199.8$~mT in steps of $1$~mT are performed. \textbf{e}, \textbf{f}, \textbf{g}, Example operation flow of the spectral-node extraction process, showing the spectral ranges of the BPFs, the corresponding filtered responses, and the detected responses, respectively.}\label{fig:hardware_demo}
\end{figure}

Guided by the above observations, we experimentally implement SDRC with a hardware-efficient filter-based design. Figure~\ref{fig:hardware_demo}a shows the experimental setup for the hardware implementation of SDRC, and Fig.~\ref{fig:hardware_demo}b illustrates its corresponding circuit diagram. The spin-wave response from each of the seven detectors is split into eight parallel channels using power splitters, with each channel filtered by a band-pass filter (BPF) covering a narrow frequency band, resulting in a total of $56$ spectral nodes. These BPFs are strategically chosen to collectively span the spectral band of $1.5$--$2.5$~GHz (Fig.~\ref{fig:hardware_demo}e), determined according to the optimal conditions observed in Figs.~\ref{fig:node_dependence}a and~\ref{fig:node_dependence}c. The filtered signals capture densely embedded spectral dynamics in their temporal responses, with examples shown in Fig.~\ref{fig:hardware_demo}f. The signal envelope of each filtered response is subsequently detected using diodes within every small window of symbol duration, forming the reservoir states for readout processing (Fig.~\ref{fig:hardware_demo}g). Power amplifiers (PAs) are inserted between splitters to ensure sufficient signal levels for detection while remaining within the linear amplification regime. Other details of hardware implementation are provided in Methods.

Figures~\ref{fig:hardware_demo}c and~\ref{fig:hardware_demo}d present the performance of the hardware-implemented SDRC system on the two benchmark tasks versus the bias field. Consistent with the emulation results, optimal performance for both tasks is achieved around a bias field of $190$~mT, where the selected BPFs effectively capture the spectral dynamics of spin waves. As the field deviates, the excited spin waves shift outside the spectral passbands of the filters, and performance degrades as a consequence. Operating the hardware-implemented SDRC system at a bias field of $188.8$~mT, we report a parity-check capacity of $3.31$, which represents state-of-the-art performance among other reported PRC systems (Fig.~\ref{fig:benchmark_performance}e). For NARMA-$2$, we obtain an NMSE of $6.8 \times 10^{-3}$ at a bias field of $174.6$~mT, comparable to state-of-the-art results obtained by PRCs with more than twice the number of nodes~(Fig.~\ref{fig:benchmark_performance}i). Notably, the performance in NARMA-$2$ is further improved compared to the emulation trends. This enhancement likely arises from differences in spectral ranges and frequency-response characteristics between the employed physical BPFs and the emulated ones, together with additional diode-induced nonlinearity and memory. These hardware-specific signatures collectively enrich the reservoir dynamics, indicating significant potential for further performance improvements through hardware optimization. We note that BPFs and diodes alone exhibit limited computational capability, as verified by omitting the spin-wave reservoir in Supplementary Fig.~2.

\section{Speech recognition}\label{sec:speech_recognition}

\begin{figure}
\centering
\includegraphics[width=1\textwidth]{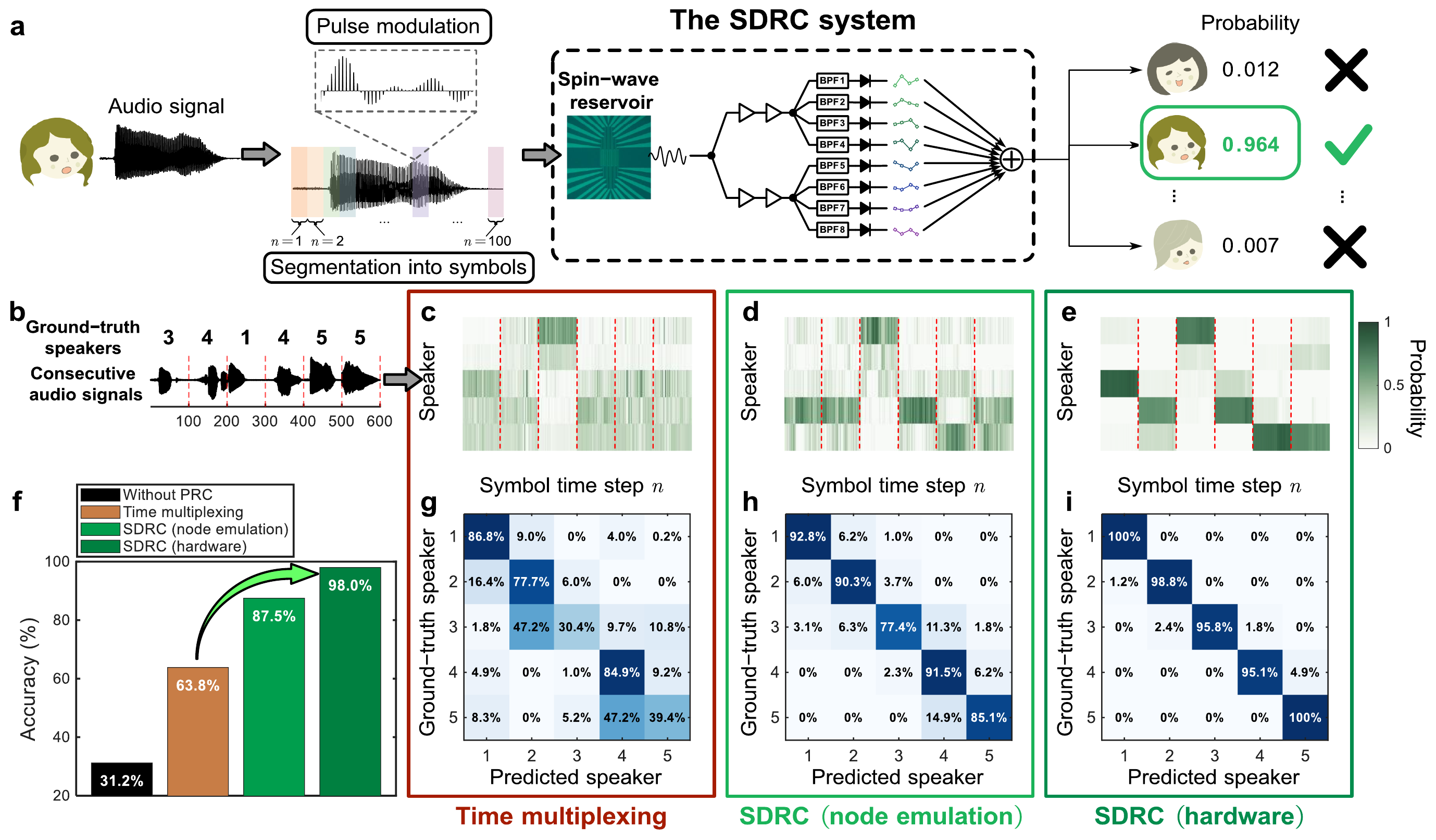}
\caption{\textbf{\textbar\ Speech recognition with the SDRC system.} \textbf{a}, Workflow of the speech-recognition process using the SDRC system. An audio signal is modulated into consecutive pulses and segmented into symbols for high-speed processing in a real-time manner. The modulated signal is then fed into the SDRC system, which outputs speaker probabilities at each symbol time step. \textbf{b}, Example waveforms of consecutive input audio signals from different speakers with their ground-truth labels. \textbf{c}, \textbf{d}, \textbf{e}, Probability colormaps for the five speakers plotted versus time step using spin-wave reservoir with emulated time multiplexing, SDRC with emulated spectral nodes, and hardware-implemented SDRC, respectively. \textbf{f}, Overall speech-recognition accuracy for the three approaches. \textbf{g}, \textbf{h}, \textbf{i}, Confusion matrices showing per-speaker prediction accuracies for spin-wave reservoir with time multiplexing, SDRC with emulated spectral nodes, and hardware-implemented SDRC, respectively.}\label{fig:speech}
\end{figure}

To demonstrate the real-world applicability of the SDRC system, we evaluate its performance on a speech-recognition task in which the system identifies the corresponding speaker given an input audio signal (Fig.~\ref{fig:speech}a). A subset of the TI-46 speech corpus dataset is used~\cite{LDC93S9}, consisting of $500$ audio recordings from five female speakers. Aiming at more real-time processing of speech signals, we avoid cochleagram or other similar feature-extraction techniques that are commonly used in PRC demonstrations~\cite{zolfagharinejad2025analogue,abreu2020role,dang2026spiking,cai2023brain}. Instead, we directly feed the time-sequential audio signals into the SDRC system after pulse modulation (Methods).  Each audio signal is segmented into $100$ temporal blocks treated as symbols, and the system outputs speaker-probability estimates at every symbol time step in a streaming manner. 

Figure~\ref{fig:speech}b illustrates an example of concatenated input audio signals together with their ground-truth speaker labels. Here, we compare the performance among time multiplexing and spectral dynamics extraction, implemented both in software (emulated) and hardware, all operating at a bias field of $187.3$~mT. Because of the ineffectiveness in information extraction, time multiplexing with a small number of nodes yields uncertain and incorrect real-time decisions throughout the recognition process (Fig.~\ref{fig:speech}c). On the other hand, SDRC effectively extracts densely encoded, speaker-specific spectral dynamics from nonlinear spin waves to produce accurate, confident, and prompt predictions (Fig.~\ref{fig:speech}d), with the hardware implementation further enhancing real-time recognition performance (Fig.~\ref{fig:speech}e). 

Figure~\ref{fig:speech}f summarizes the overall recognition accuracies for the three approaches, which are computed by assigning, for each audio signal, the speaker who receives the highest probability most often across the $100$ symbol time steps as the final prediction. Figures~\ref{fig:speech}g--i present their corresponding confusion matrices. Although time multiplexing improves accuracy ($63.8\%$) compared with direct processing without PRC ($31.2\%$), its performance remains insufficient for reliable recognition. SDRC bridges this gap by effectively extracting informative features, achieving an accuracy of $87.5\%$ in emulation and further improved to $98.0\%$ in hardware implementation.

\section{Conclusion}\label{sec:conclusion}

We have reported an SDRC system with hardware implementation that uses a small set of BPFs and diodes to extract spectral dynamics as parallel reservoir states for neuromorphic processing. Our SDRC framework exploits the oscillatory nature of spin-wave responses by extracting their coarse but rich nonlinear spectral dynamics as effective information, while reducing hardware complexity and relaxing the requirements for high-speed sampling and data handling compared with conventional information-extraction schemes based on time multiplexing. Supported by physically interpretable spectral principles that guide optimal node selection, we showed that the hardware-implemented SDRC system can achieve state-of-the-art-level performance on benchmark tasks with only a small number of nodes. Furthermore, our SDRC system demonstrates high accuracy in a real-world speech-recognition task with direct processing of time-sequential audio inputs, highlighting its strong capability for real-time computation.

Notably, our results confirmed that distinguishable and finely resolved spectral features are not mandatory for high-performance neuromorphic processing. Even with a small number of nodes concentrated within a coarsely resolved spectral range, the SDRC framework can effectively extract high-dimensional information by capturing subtle inter-modal spectral dynamics of the spin-wave responses. Moreover, our approach demonstrates that robust computational performance is realizable even at processing rates approaching the natural oscillation rate of the physical materials (with examples of faster processing rates in Supplementary Fig.~3), suggesting the potential of existing PRCs to operate at substantially higher speeds.

% SDRC differs fundamentally from frameworks that utilize well-resolved frequency spectra as reservoir states~\cite{korber2023pattern,lee2024task,heins2025benchmarking}. In those schemes, computation relies on direct access to frequency-domain representations, necessitating dedicated sampling and measurement modules, such as fast FT units or spectrum analyzers, which introduce computational overhead and operation latency that hinder true real-time processing. 

% In photonic reservoir computing, spectral (wavelength) channelization can be employed, for instance through microcombs combined with microring resonators~\cite{butschek2022photonic,lupo2023deep}. Our results demonstrate that photonic systems can potentially operate on much faster temporal scales, wherein the effective time step can approach the optical frequency range of THz. However, for these photonic systems to be scalable, it is necessary to tackle challenges such as enhancing sparse nonlinear couplings and minimizing the required implementation-heavy peripherals.

Moving forward, the SDRC framework offers a versatile paradigm for next-generation physical neuromorphic hardware. Although demonstrated here using spin-wave dynamics, the architecture of SDRC is inherently platform-agnostic and could be extended to other nonlinear physical substrates characterized by oscillatory responses, such as micro-electromechanical systems~\cite{dion2018reservoir}, superconducting circuits~\cite{govia2021quantum}, or electromagnetic cavity resonators~\cite{del2018leveraging}. From a hardware perspective, SDRC is highly amenable to architectural optimization; by strategically tuning filter center frequencies and bandwidths, the system can adaptively utilize computationally dense information subtly embedded in the spectral dynamics. Such optimization can maximize system performance under strict node-number constraints, as demonstrated by a parity-check capacity of $4.0$ with $56$ emulated nodes (Supplementary Fig.~4). Furthermore, leveraging the inherent scalability of the SDRC architecture, we expect hardware development with increased node density to give clear improvements in system performance. We have confirmed that increasing the number of nodes combined with filter tuning enables parity-check capacities exceeding $4.5$ using $140$ emulated nodes (Supplementary Fig.~4). Combined with the electronic compatibility of our approach, this architecture is well-suited for hardware-friendly integration with analogue readout circuits, such as memristive weighting networks~\cite{im2020memristive,zhong2022memristor}, paving the way toward fully end-to-end analogue pipelines for high-performance, real-time neuromorphic implementation.

\section{Methods}\label{sec:method}

\subsection{Device fabrication}
A double-side polished YIG single crystal of dimension $7 \times 7 \times 0.5$~mm\textsuperscript{3} with $(111)$-orientation was used as the spin-wave medium. The YIG sample was fabricated using the flux-growth method and supplied by TwoLeads Corporation. Ten CPWs were patterned on the YIG surface using laser lithography, consisting of a $10$-\textmu m-wide signal line and two $20$-\textmu m-wide ground lines with line gaps of $10$ \textmu m. Ti ($10$ nm)/Au ($90$ nm) thin film was deposited by electron beam evaporation for the antenna fabrication (Supplementary Fig.~5).

\subsection{Analogue pulse modulation of input data}
The discrete-time input sequence $u(n)$ of respective tasks was modulated into consecutive analogue pulses $u_p(t)$ as
\begin{equation}
u_p(t)=\sum_{n=1}^N u(n)\, p\!\left(t-nT_0\right),
\end{equation}
where $N$ is the total number of discrete-time data points, $T_0$ is the data point duration, and $p(t)$ is an impulse-like trapezoidal pulse shape described by
\begin{equation}
p(t)=
\begin{cases}
\dfrac{t}{T_{\mathrm{off}}}, 
& 0 \le t < T_{\mathrm{off}}, \\[8pt]
1, 
& T_{\mathrm{off}} \le t < T_{\mathrm{off}}+T_{\mathrm{on}}, \\[8pt]
1-\dfrac{t-T_{\mathrm{off}}-T_{\mathrm{on}}}{T_{\mathrm{off}}}, 
& T_{\mathrm{off}}+T_{\mathrm{on}} \le t < 2T_{\mathrm{off}}+T_{\mathrm{on}}, \\[6pt]
0, & \text{otherwise},
\end{cases}
\end{equation}
with $T_{\mathrm{on}}$ and $T_{\mathrm{off}}$ the flat-top duration and ramp (rise/fall) duration of the pulse, respectively. We adopted $T_0 = 5$~ns, $T_{\mathrm{on}} = 0.375$~ns, and $T_{\mathrm{off}} = 0.375$~ns in the experiment.

\subsection{Spin-wave measurement for emulation of time-multiplexing and spectral dynamics extraction}
A prepared pulse train data file encoding input information was loaded into an arbitrary waveform generator (Keysight M8915A) at a sampling rate of $16$~GHz, which delivered a $600$~mV peak-to-peak electrical signal that was subsequently amplified by $24.5$~dB using a PA (ZVA-0.5W303G) and applied to the exciter CPW. Spin-wave responses from the seven detector CPWs were amplified by $34$~dB via PAs (PE15A3269) and then recorded in an oscilloscope (Tektronix MSO68B) at a sampling rate of $12.5$~GHz. The measured spin-wave responses were averaged over $500$ measurements by the oscilloscope for noise suppression.

\subsection{Software emulation of spectral and virtual nodes}
Spectral nodes were emulated in MATLAB using time-domain spin-wave responses recorded by the oscilloscope from multiple detectors. For each response, a set of second-order Butterworth BPFs was applied with a $3$-dB passband width of $0.2$~GHz, selected to be consistent with commercially available BPFs to reflect realistic implementation conditions. Here, these BPFs have center frequencies spanning from $0.1$ to $5$~GHz in steps of $0.1$~GHz, resulting in a total of $50$ spectral nodes per detector. Discrete reservoir states were subsequently obtained by computing the root mean square values of the filtered signals over each symbol duration. 

Time multiplexing was emulated by parallelizing sampled data points within each symbol duration. For a symbol duration of $5$~ns used in the benchmark tasks, $62$ sampled data points were available as virtual nodes, as determined by the oscilloscope sampling rate of $12.5$~GHz. 

For the random node-selection analysis shown in Figs.~\ref{fig:benchmark_performance}d and~\ref{fig:benchmark_performance}h, $N_{\mathrm{node}}$ nodes per detector were randomly selected from the pools of $50$ spectral nodes and $62$ virtual nodes, respectively, with $N_{\mathrm{node}} \in \{ 5, 10, 15, 20, 25, 30 \}$. This random selection procedure was repeated for $2\times10^6$ trials for each task and each $N_{\mathrm{node}}$ (except for $N_{\mathrm{node}} = 5$, for which we did all possible node combinations), and their corresponding minimum and maximum capacities and NMSEs were reported (Supplementary Note 2).

\subsection{Hardware implementation of SDRC}
The hardware implementation of SDRC follows the diagram shown in Fig.~\ref{fig:benchmark_performance}b. First, a $1$:$2$ splitter was used to split the spin-wave response signal measured from a detector antenna into two channels. Each split signal was subsequently amplified by $24.5$~dB and $20.7$~dB via two PAs (PE15A3269 and ZX60-83MP) operating within their linear regimes. Two $1$:$4$ splitters then split the amplified signals into eight channels. The eight BPFs employed in the implementation were VBF-1525+ ($1480$--$1570$~MHz), VBF-1575+ ($1530$--$1620$~MHz), VBF-1840+ ($1750$--$1930$~MHz), VBF-1945+ ($1850$--$2040$~MHz), VBF-2130-S+ ($2000$--$2260$~MHz), VBF-2275+ ($2170$--$2380$~MHz), VBF-2360+ ($2250$--$2470$~MHz), and VBF-2435+ ($2340$--$2530$~MHz). Eight diodes (CPDETLS-4000) were used to rectify the filtered responses. An oscilloscope (Tektronix MSO68B) at a sampling rate of $12.5$~GHz was used to measure the rectified responses for all seven detector CPWs by sequentially switching the terminal connections. During these measurements, any detector terminal not actively connected to the oscilloscope was terminated with a $50\ \Omega$ attenuator to maintain impedance matching. Finally, the mean value of recorded states within each symbol duration was calculated in post-processing. In a practical device, this process can be straightforwardly realized by sampling the rectified signals at a slower clock rate matched to the symbol duration.

\subsection{Speech-recognition task}

The speech-recognition task was performed using $500$ audio recordings from the TI-46 speech corpus dataset. These recordings were taken from five female speakers, each speaker uttering the ten spoken digits ten times, yielding 100 samples per speaker. All audio signals were standardized to a uniform length of $10000$ data points by partially trimming silent segments at the beginning and end. Each audio signal was segmented into $100$ symbols for real-time processing. The target signal was defined as the ground-truth probability array $\hat{\mathbf{y}}(n) = [y_1(n), y_2(n),y_3(n),y_4(n),y_5(n)]$, with $y_i(n) = 1$ for the index corresponding to the true speaker and $0$ otherwise. Here, for each audio sample, $n$ denotes the symbol time step ranging from $1$ to $100$. To ensure statistical robustness, the $500$ audio samples were randomly shuffled twenty times; in each trial, $400$ were used for training and the remaining $100$ for testing. Other details about readout training and testing algorithms are provided in Supplementary Note~2. A winner-takes-all rule was applied at each symbol time step to determine the instantaneous prediction, and the speaker index occurring most often across the $100$ steps of a given audio signal was taken as the final decision. For evaluation, we compared the final decision of each audio signal with its ground-truth label to calculate the recognition accuracy, which was then averaged over the twenty trials.

\bmhead{Acknowledgements}

This work was in part supported by Innovative Science and Technology Initiative for Security Grant Number JPJ004596, ATLA, Japan, KIOXIA Corporation, and the Royal Society International Collaboration Award (Grant Number ICA\textbackslash R1\textbackslash231091). A part of this work was supported by “Advanced Research Infrastructure for Materials and Nanotechnology in Japan (ARIM)" of the Ministry of Education, Culture, Sports, Science and Technology (MEXT). Proposal Number JPMXP1225NM5244, 25NM5247 and 25NM5248.

\bmhead{Competing interests}
The authors declare the following competing interests: A Japanese patent application related to the spectral dynamics reservoir computing framework described in this manuscript has been filed by the National Institute for Materials Science (NIMS). The inventors are Jiaxuan Chen, Ryo Iguchi, and Takashi Tsuchiya. The application number is JP 2026‑013405 (filed 29 January 2026), and it is currently pending. The patent application covers aspects of spectral dynamics reservoir computing using analogue filtering and envelope detection for neuromorphic processing.

%%===========================================================================================%%
%% If you are submitting to one of the Nature Portfolio journals, using the eJP submission   %%
%% system, please include the references within the manuscript file itself. You may do this  %%
%% by copying the reference list from your .bbl file, paste it into the main manuscript .tex %%
%% file, and delete the associated \verb+\bibliography+ commands.                            %%
%%===========================================================================================%%

% \bibliography{sn-bibliography}% common bib file
%% if required, the content of .bbl file can be included here once bbl is generated
\input Main.bbl

\clearpage


\section*{Supplementary material}

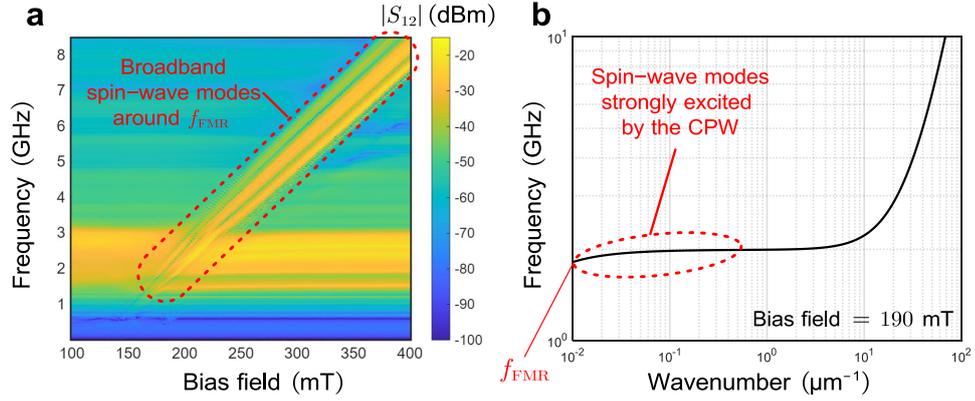

**Supplementary Fig. 1 | Spin wave transmission spectroscopy and approximated dispersion relation. a**, Transmission spectrum obtained via vector network analyzer (VNA) measurements, showing the magnitude of the scattering parameter $|S_{12}|$ in dBm with respect to the magnetic bias field and excitation frequency. For this measurement, the fourth coplanar waveguide (CPW) from the left was used as the exciter and the third CPW from the right as the detector (Supplementary Fig. 5). We observe that multiple branches of broad spin-wave modes were excited around the ferromagnetic resonance (FMR) frequency, likely attributed to the large film thickness that supports a variety of spin-wave thickness modes. We also note that strong electromagnetic induction was observed around 2 GHz. These spectral features correlate with the node distributions in Fig. 4 of the main text, providing a physically grounded interpretation of the spectral principles for information-effective node optimization. **b**, Approximated dispersion relation of the lowest spin-wave branch between frequency and wavenumber. The red circle indicates a rough estimation of the strongly excited spin-wave modes based on the CPW geometry, which corresponds to the observed dominant spin-wave modes in the VNA measurements. This relation was estimated following equation (S9), with parameters $B_0 = 190$ mT, $M_s = 130$ kA/m, $D = 5.5 \times 10^{-17}$ Tm$^2$, and $d = 0.5$ mm.

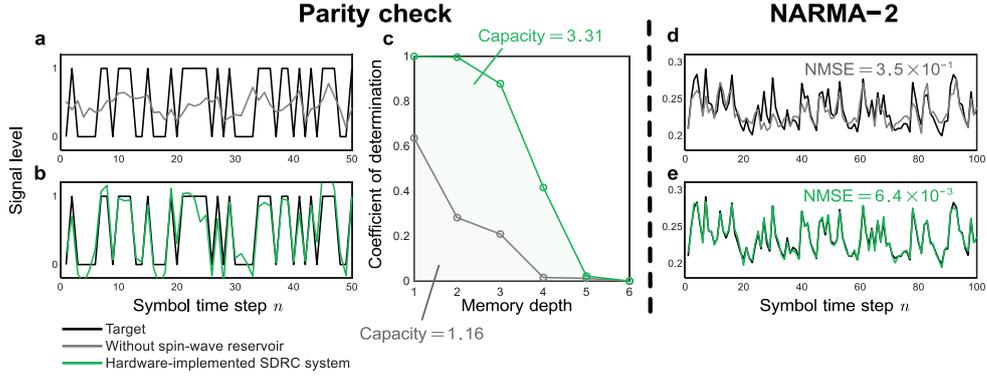

**Supplementary Fig. 2 | Contribution of the spin-wave reservoir to the computational performance of spectral dynamics reservoir computing (SDRC).** We compare the performance of the full hardware-implemented SDRC system against a baseline configuration excluding the spin-wave reservoir. This comparison clarifies the contribution of the spin-wave reservoir and demonstrates the computational capability of the spectral dynamics extraction layer, involving only band-pass filters (BPFs) and diodes. **a,b**, Example time-domain waveforms for the parity-check task at a memory depth $K = 3$, showing the target signal (black) and the predicted signals obtained without spin-wave reservoir (gray) and with the full hardware-implemented SDRC system (green). We observed that BPFs and diodes alone exhibited limited computational abilities. **c**, Corresponding coefficients of determination plotted against memory depth $K$, where the parity-check capacity decreases from 3.31 for the full system to 1.16 for only BPFs and diodes. **e**, **f**, Example time-domain waveforms for the NARMA-2 task, showing the target signal (black) and the predicted signals obtained without spin-wave reservoir (gray) and with the full hardware-implemented SDRC system (green), with the normalized mean square error (NMSE) degrades significantly from $6.4 \times 10^{-3}$ with the full SDRC system to $3.5 \times 10^{-1}$ with only BPFs and diodes. This performance gap confirms that the BPFs and diodes alone provide insufficient dimensionality for complex tasks, underscoring that the system's high computational power mainly originates from the high-dimensional, nonlinear dynamics attributed to the spin-wave reservoir.

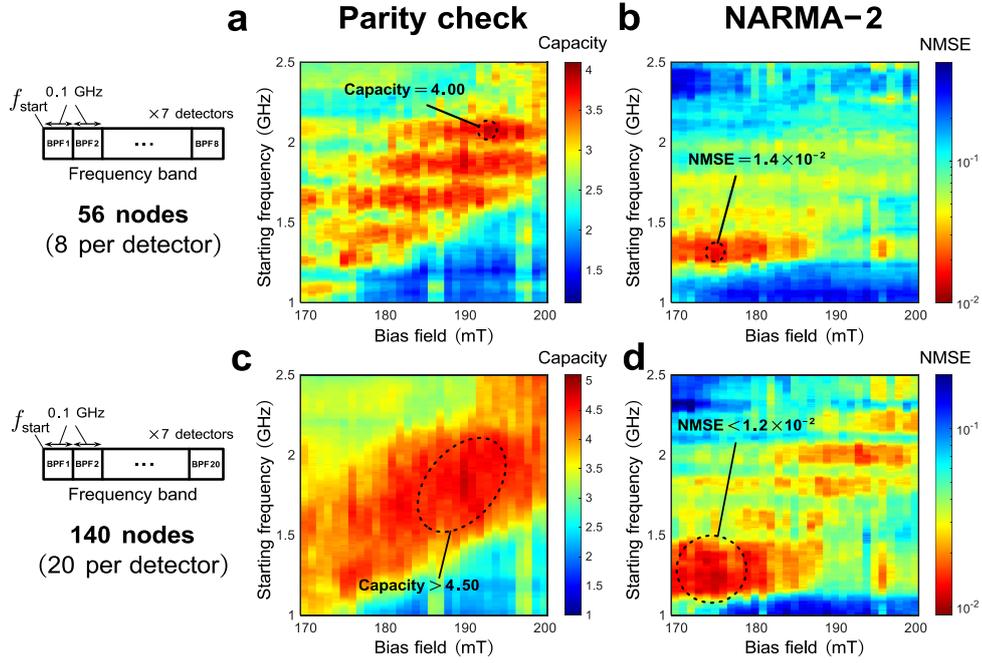

Supplementary Fig. 3 | **Performance optimization with fine filter tuning.** To demonstrate the potential of adaptability and scalability of the SDRC framework, we emulated the spectral dynamics extraction process across refined filter configurations for both benchmark tasks. Specifically, the bandwidth of the BPFs was narrowed from 0.2 to 0.1 GHz. These filters were distributed across the spectral manifold starting from an initial frequency $f_{\text{start}}$, with their center frequencies uniformly spaced at fine intervals of 0.01 GHz. We swept the starting frequency $f_{\text{start}}$ from 1 to 2.5 GHz in steps of 5 MHz and the bias field from 169.8 to 199.8 mT in steps of approximately 1 mT to identify optimal operating regimes. **a**, Optimized parity-check capacity with the node number fixed at 56 (8 per detector) across various bias fields and filter's starting frequencies. Fine tuning of the filter distribution enhances capacity from 3.51 (when using spectral nodes across the full spectrum with 50 nodes per detector, as shown in Fig. 3 of the main text) to a peak value of 4.00. **b**, Optimized NMSE for the NARMA-2 task with the node number fixed at 56 (8 per detector), showing an improvement in NMSE from $1.9 \times 10^{-2}$ (Fig. 3 of the main text) to a minimum value of $1.4 \times 10^{-2}$ with fine filter tuning. **c**, Further optimization of parity-check capacity when increasing the node number to 140 (20 per detector), where a variety of combinations of starting frequencies $f_{\text{start}}$ and bias fields give capacities exceeding 4.5. **d**, Futher optimization of NARMA-2 NMSE when increasing the node number to 140 (20 per detector), where a variety of combinations of starting frequencies $f_{\text{start}}$ and bias fields give NMSE below $1.2 \times 10^{-2}$. We note that these performance can be further improved through more careful optimization of filtering choices, such as densely distributing filters' center frequencies across multiple spectral zones.

4

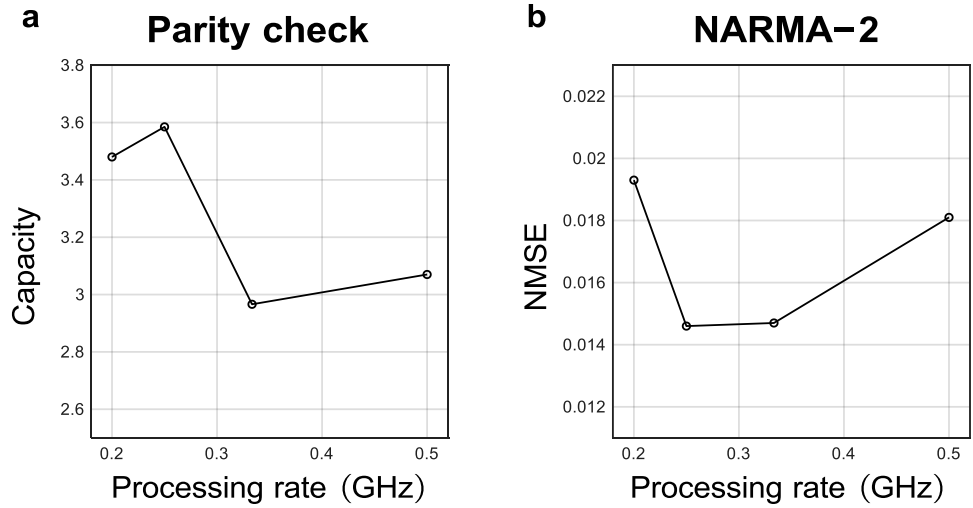

**Supplementary Fig. 4 | Computational performance of SDRC at higher processing rates.** We evaluated the computational robustness of the SDRC framework at processing rates approaching the material's intrinsic dynamical timescales. We examined SDRC with emulated spectral nodes in the two benchmark tasks at different processing speeds, with experiments conducted at a bias field of 187.3 mT, corresponding to a dominant spin-wave resonance frequency around 2 GHz. The processing rate is defined as the inverse of the symbol duration, as described in Supplementary Note 4. **a**, **b**, The parity-check capacity and NARMA-2 NMSE plotted against the processing rate, respectively. Beyond the processing rate of 0.2 GHz, as we adopted in the primary experiments (main text), SDRC shows robust performance even at higher rates, such as a capacity of 3.07 and an NMSE of $1.8 \times 10^{-2}$ at a processing rate of 0.5 GHz. This processing rate is very close to the natural oscillation frequency of spin waves, confirming its potential for high-speed neuromorphic computing.

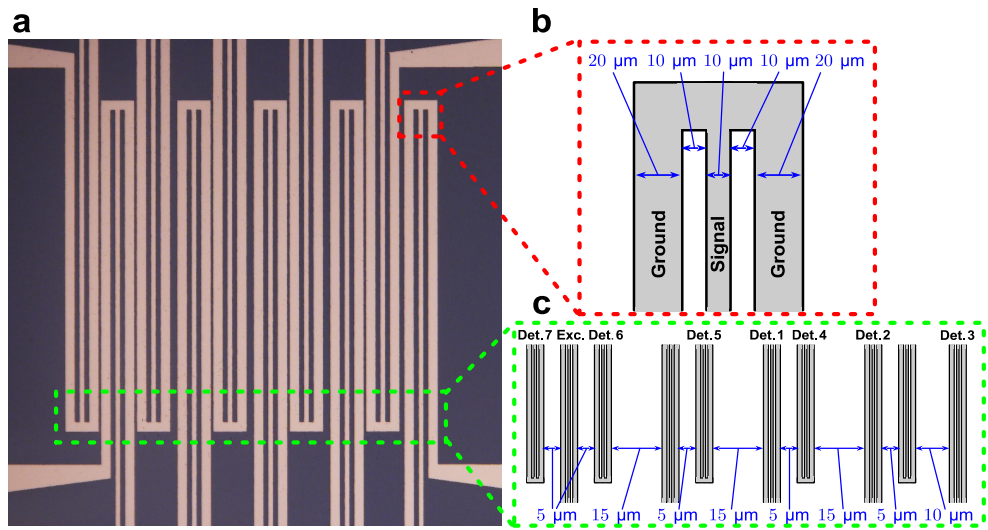

**Supplementary Fig. 5 | Antenna structure. a**, An optical microscope image showing the structure of ten CPWs fabricated on an yttrium iron garnet (YIG) medium. **b**, Geometric specifications of each CPW antenna, including the lengths of the ground and signal lines and the gap spacing. **c**, Overall antenna layout with inter-antenna separations indicated, where one CPW is used as the spin-wave exciter and seven as detectors.

6

# Supplementary Note 1. Theoretical background of spin-wave nonlinear dynamics

Spin-wave dynamics are theoretically governed by the Landau-Lifshitz-Gilbert (LLG) equation:

$$\frac{\partial \mathbf{M}}{\partial t} = -\gamma\, \mathbf{M} \times \mathbf{B}_{\text{eff}} + \frac{\alpha}{M_{\text{s}}} \mathbf{M} \times \frac{\partial \mathbf{M}}{\partial t}, \tag{S1}$$

where $\mathbf{M}$ is the magnetization vector, $M_{\text{s}}$ is the saturation magnetization of the material, $\mathbf{B}_{\text{eff}}$ is the effective magnetic field, $\gamma$ is the gyromagnetic ratio, $\mu_0$ is the vacuum permeability, and $\alpha$ is the Gilbert damping constant [1–3]. Typically, $\mathbf{B}_{\text{eff}}$ includes

$$\mathbf{B}_{\text{eff}} = \mathbf{B}_{\text{a}} + \mathbf{B}_{\text{d}} + \mathbf{B}_{\text{ex}}, \tag{S2}$$

where $\mathbf{B}_{\text{a}}$ is the externally applied magnetic field, $\mathbf{B}_{\text{d}}$ is the demagnetizing field arising from the magnetic dipole interactions, and $\mathbf{B}_{\text{ex}}$ is an effective field accounting for exchange interactions. Owing to the nonlinear nature of the LLG equation, spin waves exhibit rich dynamics with both oscillatory and relaxational characteristics, arising from the precessional and damping terms, respectively.

In the linear-regime approximation of weak excitation, spin waves of different frequencies propagate independently and can be treated as distinct dynamical modes. Therefore, we can approximate their collections as a superposition of complex exponentials, expressed as

$$\mathbf{M}(\mathbf{r}, t) = \sum_n \mathbf{M}_n e^{i(\mathbf{k}_n \cdot \mathbf{r} - \omega_n t)}, \tag{S3}$$

with $\mathbf{M}_n$, $\mathbf{k}_n$, and $\omega_n$ the amplitude vector, wavevector, and angular frequency of the $n$-th spin-wave mode, $\mathbf{r}$ the position vector, and $i$ the imaginary unit. Moreover, to simplify the expression for $\mathbf{B}_{\text{eff}}$, we neglect the excitation field and assume

$$\mathbf{B}_{\text{a}}(\mathbf{r}, t) = \mathbf{B}_0, \tag{S4}$$

with $\mathbf{B}_0$ the spatially uniform, time-independent bias magnetic field for spin-wave stabilization. For the demagnetization field $\mathbf{B}_{\text{d}}$, we invoke the Green functions to express them as

$$\mathbf{B}_{\text{d}}(\mathbf{r}, t) = \int d^3 r'\, \mathbf{G}(\mathbf{r} - \mathbf{r}')\, \mathbf{M}(\mathbf{r}', t), \tag{S5}$$

where $\mathbf{G}(\mathbf{r})$ is the magnetostatic Green's tensor describing the spatially nonlocal dipolar interaction. The exchange field $\mathbf{B}_{\text{d}}$ is expressed as

$$\mathbf{H}_{\text{ex}}(\mathbf{r}, t) = D\, \nabla^2 \mathbf{M}(\mathbf{r}, t), \tag{S6}$$

with $\nabla^2$ the Laplacian operator and $D$ the exchange stiffness parameter. Substituting equation (S3) into equation (S2) with the above expressions for individual field terms,

we obtain

$$\begin{aligned}\mathbf{H}_{\text{eff}}(\mathbf{r},t) &= \mathbf{B}_0 + \sum_n \hat{\mathbf{G}}(\mathbf{k}_n)\mathbf{M}_n e^{i(\mathbf{k}_n\cdot\mathbf{r}-\omega_n t)} - \sum_n D|\mathbf{k}_n|^2 \mathbf{M}_n e^{i(\mathbf{k}_n\cdot\mathbf{r}-\omega_n t)} \\ &= \mathbf{B}_0 + \sum_n \hat{\mathcal{G}}(\mathbf{k}_n)\mathbf{M}_n e^{i(\mathbf{k}_n\cdot\mathbf{r}-\omega_n t)},\end{aligned} \quad (S7)$$

where $\hat{\mathbf{G}}(\mathbf{k})$ is the dipolar Green's tensor in reciprocal space, expressed as

$$\hat{\mathbf{G}}(\mathbf{k}) = \int d^3\rho \, \mathbf{G}(\rho) \, e^{i\mathbf{k}\cdot\rho}, \quad (S8)$$

and $\hat{\mathcal{G}}(\mathbf{k}) = \hat{\mathbf{G}}(\mathbf{k}) - D|\mathbf{k}_n|^2$ is the combined dipolar-exchange Green's tensor.

By substituting equations (S3) and (S7) into the LLG equation (S1) and keeping only the linear-order terms, we obtain the dispersion relations that characterize the dynamics of individual spin-wave modes. Specifically, for forward volume spin waves (FVSWs), where the magnetization is oriented out-of-plane, the dispersion relation is given by

$$f(k) = \frac{\gamma}{2\pi}\sqrt{(B_0 - \mu_0 M_{\text{s}} + Dk^2)(B_0 - \mu_0 M_{\text{s}} + Dk^2 + \mu_0 M_{\text{s}} F(kd))}, \quad (S9)$$

where $k$ and $f$ are the wavenumber and corresponding frequency of the spin-wave mode, $B_0$ is the out-of-plane bias field strength, $d$ is the film thickness, and $F(kd)$ represents the dipole-dipole form factor. As a consequence of this nonlinear dispersion, spin-wave modes at different frequencies possess distinct group velocities and precession amplitudes in propagation. Spin waves exhibit strong resonance characteristics near the long-wavelength limit ($k \to 0$), known as FMR. For the FVSW geometry, the FMR frequency is expressed by the Kittel formula as

$$f_{\text{FMR}} = \frac{\gamma}{2\pi}(B_0 - \mu_0 M_{\text{s}}). \quad (S10)$$

This frequency depends on the strength of the bias magnetic field $B_0$ and defines the lower bound of the spin-wave spectrum.

Beyond the linear regime, the nonlinear terms in the LLG equation (S1) must be taken into account. Consequently, we obtain the expression characterizing nonlinear inter-modal interactions of spin waves:

$$\sum_n i\omega_n \mathbf{M}_n e^{i(\mathbf{k}_n\cdot\mathbf{r}-\omega_n t)} = \gamma \left(\sum_n \mathbf{M}_n e^{i(\mathbf{k}_n\cdot\mathbf{r}-\omega_n t)}\right) \times \left(\sum_n \hat{\mathcal{G}}(\mathbf{k}_n)\mathbf{M}_n e^{i(\mathbf{k}_n\cdot\mathbf{r}-\omega_n t)} + \mathbf{B}_0\right). \quad (S11)$$

Here, each complex exponential of a particular mode component $\omega_n$ needs to satisfy the equality on both sides, and consequently the cross product generates higher-order mixing terms of the form $e^{i((\mathbf{k}_{n_1}\pm\mathbf{k}_{n_2})\cdot\mathbf{r}-(\omega_{n_1}\pm\omega_{n_2})t)}$. This indicates that the spin-wave modes are nonlinearly coupled, giving rise to the latent dense spectral manifold through complex frequency-mixing processes.

# Supplementary Note 2. Task specifications and reservoir computing algorithms

## Parity check

The target signal $\hat{y}(n)$ of the parity-check task is expressed as

$$\hat{y}(n) = \bigoplus_{i=0}^{K-1} u(n-i), \tag{S12}$$

where $\bigoplus$ represents the addition modulo 2 (i.e., exclusive OR), $u(n)$ is a binary input sequence uniformly randomly sampled from $\{0, 1\}$, and $K$ is the memory depth. A total of $N = 2700$ steps of $u(n)$ were prepared, with steps in $1 \leq n \leq 200$ discarded to avoid transient effects, $201 \leq n \leq 2200$ used for the training phase, and $2201 \leq n \leq 2700$ used for the testing phase. Performance at a specific memory depth $K$ was evaluated by calculating the coefficient of determination $R_K^2$ between the predicted output $y_K(n)$ and the target output $\hat{y}_K(n)$ during the testing phase, expressed as

$$R_K^2 = 1 - \frac{\sum_n (y_K(n) - \hat{y}_K(n))^2}{\sum_n (\hat{y}_K(n) - \langle \hat{y}_K \rangle)^2}, \tag{S13}$$

with $\langle \hat{y}_K \rangle$ denoting the average of $\hat{y}_K(n)$. The parity-check capacity is then calculated by

$$C_{\text{total}} = \sum_{K=1}^{K_0} R_K^2, \tag{S14}$$

with $K_0 = 6$ in this study.

## NARMA-2

The target signal $\hat{y}(n)$ of the NARMA-2 tasks is described by the following recursive relation:

$$\hat{y}(n+1) = 0.4\,\hat{y}(n) + 0.4\,\hat{y}(n)\,\hat{y}(n-1) + 0.6\,u(n)^3 + 0.1, \tag{S15}$$

where $u(n)$ is an input sequence randomly sampled uniformly from $[0, 0.5]$. Again, an input signal $u(n)$ of length $N = 2700$ was used, with $1 \leq n \leq 200$ discarded, $201 \leq n \leq 2200$ for training, and $2201 \leq n \leq 2700$ for testing. Performance was quantified by calculating the NMSE between predicted output $y_K(n)$ and target output $\hat{y}_K(n)$ during the testing phase, expressed as

$$\text{NMSE} = \frac{\sum_n (y(n) - \hat{y}(n))^2}{\sum_n (\hat{y}(n) - \langle \hat{y} \rangle)^2}, \tag{S16}$$

## Readout training and testing

During the training phase, the readout weight matrix $\mathbf{W} = [w_{i,j}] \in \mathbb{R}^{I \times O}$ was trained by linear regression as

$$\mathbf{W} = (\mathbf{X}^{\mathrm{T}}\mathbf{X})^{-1}\mathbf{X}^{\mathrm{T}}\hat{\mathbf{Y}}. \tag{S17}$$

Here, $\mathbf{X} \in \mathbb{R}^{N_s \times I}$ is the reservoir state matrix used for training, with each node column-wisely aligned, $\hat{\mathbf{Y}} \in \mathbb{R}^{N_s \times O}$ is the target output matrix, $(\cdot)^{\mathrm{T}}$ represents the matrix transposition, $I$ is the number of reservoir states (nodes), $O$ is the number of target output channels ($O = 1$ for parity check and NARMA-2 and $O = 5$ for speech recognition), and $N_s$ is the number of symbols. We note that in the parity-check and NARMA-2 tasks, one data point corresponds to one symbol, i.e., $N_s = N$. In the speech-recognition task, we treated 100 data points as one symbol, i.e., $N_s = N/100$. Accordingly, the symbol duration is $T_s = T_0$ for the parity-check and NARMA-2 tasks, and $T_s = 100T_0$ for the speech-recognition task. The processing rate is calculated as the inverse of $T_s$.

During the testing phase, the output matrix $\mathbf{Y} \in \mathbb{R}^{N_s \times O}$ is generated by a linear combination of the reservoir states with trained weights, expressed as

$$\mathbf{Y} = \mathbf{X}\mathbf{W}. \tag{S18}$$

# Supplementary References

[1] Stancil, D.D., Prabhakar, A.: Spin Waves: Theory and Applications. Springer, New York, NY (2009)

[2] Gurevich, A.G., Melkov, G.A.: Magnetization Oscillations and Waves. CRC Press, Boca Raton, FL (1996)

[3] Mahmoud, A., Ciubotaru, F., Vanderveken, F., Chumak, A.V., Hamdioui, S., Adelmann, C., Cotofana, S.: Introduction to spin wave computing. J. Appl. Phys. **128**, 161101 (2020)



\end{document}

%% file: Main.bbl
%% BioMed_Central_Bib_Style_v1.01

%% file: Main.bbl
\begin{thebibliography}{64}
% BibTex style file: bmc-mathphys.bst (version 2.1), 2014-07-24
\ifx \bisbn   \undefined \def \bisbn  #1{ISBN #1}\fi
\ifx \binits  \undefined \def \binits#1{#1}\fi
\ifx \bauthor  \undefined \def \bauthor#1{#1}\fi
\ifx \batitle  \undefined \def \batitle#1{#1}\fi
\ifx \bjtitle  \undefined \def \bjtitle#1{#1}\fi
\ifx \bvolume  \undefined \def \bvolume#1{\textbf{#1}}\fi
\ifx \byear  \undefined \def \byear#1{#1}\fi
\ifx \bissue  \undefined \def \bissue#1{#1}\fi
\ifx \bfpage  \undefined \def \bfpage#1{#1}\fi
\ifx \blpage  \undefined \def \blpage #1{#1}\fi
\ifx \burl  \undefined \def \burl#1{\textsf{#1}}\fi
\ifx \doiurl  \undefined \def \doiurl#1{\url{https://doi.org/#1}}\fi
\ifx \betal  \undefined \def \betal{\textit{et al.}}\fi
\ifx \binstitute  \undefined \def \binstitute#1{#1}\fi
\ifx \binstitutionaled  \undefined \def \binstitutionaled#1{#1}\fi
\ifx \bctitle  \undefined \def \bctitle#1{#1}\fi
\ifx \beditor  \undefined \def \beditor#1{#1}\fi
\ifx \bpublisher  \undefined \def \bpublisher#1{#1}\fi
\ifx \bbtitle  \undefined \def \bbtitle#1{#1}\fi
\ifx \bedition  \undefined \def \bedition#1{#1}\fi
\ifx \bseriesno  \undefined \def \bseriesno#1{#1}\fi
\ifx \blocation  \undefined \def \blocation#1{#1}\fi
\ifx \bsertitle  \undefined \def \bsertitle#1{#1}\fi
\ifx \bsnm \undefined \def \bsnm#1{#1}\fi
\ifx \bsuffix \undefined \def \bsuffix#1{#1}\fi
\ifx \bparticle \undefined \def \bparticle#1{#1}\fi
\ifx \barticle \undefined \def \barticle#1{#1}\fi
\bibcommenthead
\ifx \bconfdate \undefined \def \bconfdate #1{#1}\fi
\ifx \botherref \undefined \def \botherref #1{#1}\fi
\ifx \url \undefined \def \url#1{\textsf{#1}}\fi
\ifx \bchapter \undefined \def \bchapter#1{#1}\fi
\ifx \bbook \undefined \def \bbook#1{#1}\fi
\ifx \bcomment \undefined \def \bcomment#1{#1}\fi
\ifx \oauthor \undefined \def \oauthor#1{#1}\fi
\ifx \citeauthoryear \undefined \def \citeauthoryear#1{#1}\fi
\ifx \endbibitem  \undefined \def \endbibitem {}\fi
\ifx \bconflocation  \undefined \def \bconflocation#1{#1}\fi
\ifx \arxivurl  \undefined \def \arxivurl#1{\textsf{#1}}\fi
\csname PreBibitemsHook\endcsname

%%% 1
\bibitem[\protect\citeauthoryear{Wright et~al.}{2022}]{wright2022deep}
\begin{barticle}
\bauthor{\bsnm{Wright}, \binits{L.G.}},
\bauthor{\bsnm{Onodera}, \binits{T.}},
\bauthor{\bsnm{Stein}, \binits{M.M.}},
\bauthor{\bsnm{Wang}, \binits{T.}},
\bauthor{\bsnm{Schachter}, \binits{D.T.}},
\bauthor{\bsnm{Hu}, \binits{Z.}},
\bauthor{\bsnm{McMahon}, \binits{P.L.}}:
\batitle{Deep physical neural networks trained with backpropagation}.
\bjtitle{Nature}
\bvolume{601},
\bfpage{549}--\blpage{555}
(\byear{2022})
\end{barticle}
\endbibitem

%%% 2
\bibitem[\protect\citeauthoryear{Shastri et~al.}{2021}]{shastri2021photonics}
\begin{barticle}
\bauthor{\bsnm{Shastri}, \binits{B.J.}},
\bauthor{\bsnm{Tait}, \binits{A.N.}},
\bauthor{\bsnm{Lima}, \binits{T.}},
\bauthor{\bsnm{Pernice}, \binits{W.H.P.}},
\bauthor{\bsnm{Bhaskaran}, \binits{H.}},
\bauthor{\bsnm{Wright}, \binits{C.D.}},
\bauthor{\bsnm{Prucnal}, \binits{P.R.}}:
\batitle{Photonics for artificial intelligence and neuromorphic computing}.
\bjtitle{Nat. Photonics}
\bvolume{15},
\bfpage{102}--\blpage{114}
(\byear{2021})
\end{barticle}
\endbibitem

%%% 3
\bibitem[\protect\citeauthoryear{Grollier et~al.}{2020}]{grollier2020neuromorphic}
\begin{barticle}
\bauthor{\bsnm{Grollier}, \binits{J.}},
\bauthor{\bsnm{Querlioz}, \binits{D.}},
\bauthor{\bsnm{Camsari}, \binits{K.Y.}},
\bauthor{\bsnm{Everschor-Sitte}, \binits{K.}},
\bauthor{\bsnm{Fukami}, \binits{S.}},
\bauthor{\bsnm{Stiles}, \binits{M.D.}}:
\batitle{Neuromorphic spintronics}.
\bjtitle{Nat. Electron.}
\bvolume{3},
\bfpage{360}--\blpage{370}
(\byear{2020})
\end{barticle}
\endbibitem

%%% 4
\bibitem[\protect\citeauthoryear{Roy et~al.}{2019}]{roy2019towards}
\begin{barticle}
\bauthor{\bsnm{Roy}, \binits{K.}},
\bauthor{\bsnm{Jaiswal}, \binits{A.}},
\bauthor{\bsnm{Panda}, \binits{P.}}:
\batitle{Towards spike-based machine intelligence with neuromorphic computing}.
\bjtitle{Nature}
\bvolume{575},
\bfpage{607}--\blpage{617}
(\byear{2019})
\end{barticle}
\endbibitem

%%% 5
\bibitem[\protect\citeauthoryear{Yan et~al.}{2024}]{yan2024emerging}
\begin{barticle}
\bauthor{\bsnm{Yan}, \binits{M.}},
\bauthor{\bsnm{Huang}, \binits{C.}},
\bauthor{\bsnm{Bienstman}, \binits{P.}},
\bauthor{\bsnm{Tino}, \binits{P.}},
\bauthor{\bsnm{Lin}, \binits{W.}},
\bauthor{\bsnm{Sun}, \binits{J.}}:
\batitle{Emerging opportunities and challenges for the future of reservoir computing}.
\bjtitle{Nat. Commun.}
\bvolume{15},
\bfpage{2056}
(\byear{2024})
\end{barticle}
\endbibitem

%%% 6
\bibitem[\protect\citeauthoryear{Tanaka et~al.}{2019}]{tanaka2019recent}
\begin{barticle}
\bauthor{\bsnm{Tanaka}, \binits{G.}},
\bauthor{\bsnm{Yamane}, \binits{T.}},
\bauthor{\bsnm{H{\'e}roux}, \binits{J.B.}},
\bauthor{\bsnm{Nakane}, \binits{R.}},
\bauthor{\bsnm{Kanazawa}, \binits{N.}},
\bauthor{\bsnm{Takeda}, \binits{S.}},
\bauthor{\bsnm{Numata}, \binits{H.}},
\bauthor{\bsnm{Nakano}, \binits{D.}},
\bauthor{\bsnm{Hirose}, \binits{A.}}:
\batitle{Recent advances in physical reservoir computing: a review}.
\bjtitle{Neural Netw.}
\bvolume{115},
\bfpage{100}--\blpage{123}
(\byear{2019})
\end{barticle}
\endbibitem

%%% 7
\bibitem[\protect\citeauthoryear{Appeltant et~al.}{2011}]{appeltant2011information}
\begin{barticle}
\bauthor{\bsnm{Appeltant}, \binits{L.}},
\bauthor{\bsnm{Soriano}, \binits{M.C.}},
\bauthor{\bsnm{Sande}, \binits{G.}},
\bauthor{\bsnm{Danckaert}, \binits{J.}},
\bauthor{\bsnm{Massar}, \binits{S.}},
\bauthor{\bsnm{Dambre}, \binits{J.}},
\bauthor{\bsnm{Schrauwen}, \binits{B.}},
\bauthor{\bsnm{Mirasso}, \binits{C.R.}},
\bauthor{\bsnm{Fischer}, \binits{I.}}:
\batitle{Information processing using a single dynamical node as complex system}.
\bjtitle{Nat. Commun.}
\bvolume{2},
\bfpage{468}
(\byear{2011})
\end{barticle}
\endbibitem

%%% 8
\bibitem[\protect\citeauthoryear{Jaeger}{2001}]{jaeger2001echo}
\begin{botherref}
\oauthor{\bsnm{Jaeger}, \binits{H.}}:
The ``echo state'' approach to analysing and training recurrent neural networks -- with an erratum note.
Technical Report 148,
German National Research Center for Information Technology,
Sankt Augustin, Germany
(2001)
\end{botherref}
\endbibitem

%%% 9
\bibitem[\protect\citeauthoryear{Markovi{\'c} et~al.}{2020}]{markovic2020physics}
\begin{barticle}
\bauthor{\bsnm{Markovi{\'c}}, \binits{D.}},
\bauthor{\bsnm{Mizrahi}, \binits{A.}},
\bauthor{\bsnm{Querlioz}, \binits{D.}},
\bauthor{\bsnm{Grollier}, \binits{J.}}:
\batitle{Physics for neuromorphic computing}.
\bjtitle{Nat. Rev. Phys.}
\bvolume{2},
\bfpage{499}--\blpage{510}
(\byear{2020})
\end{barticle}
\endbibitem

%%% 10
\bibitem[\protect\citeauthoryear{Zidan et~al.}{2018}]{zidan2018future}
\begin{barticle}
\bauthor{\bsnm{Zidan}, \binits{M.A.}},
\bauthor{\bsnm{Strachan}, \binits{J.P.}},
\bauthor{\bsnm{Lu}, \binits{W.D.}}:
\batitle{The future of electronics based on memristive systems}.
\bjtitle{Nat. Electron.}
\bvolume{1},
\bfpage{22}--\blpage{29}
(\byear{2018})
\end{barticle}
\endbibitem

%%% 11
\bibitem[\protect\citeauthoryear{Aadhi et~al.}{2025}]{aadhi2025scalable}
\begin{botherref}
\oauthor{\bsnm{Aadhi}, \binits{A.}},
\oauthor{\bsnm{Di~Lauro}, \binits{L.}},
\oauthor{\bsnm{Fischer}, \binits{B.}},
\oauthor{\bsnm{Dmitriev}, \binits{P.}},
\oauthor{\bsnm{Alamgir}, \binits{I.}},
\oauthor{\bsnm{Mazoukh}, \binits{C.}},
\oauthor{\bsnm{Perron}, \binits{N.}},
\oauthor{\bsnm{Viktorov}, \binits{E.A.}},
\oauthor{\bsnm{Kovalev}, \binits{A.V.}},
\oauthor{\bsnm{Eshaghi}, \binits{A.}}, et al.:
Scalable photonic reservoir computing for parallel machine learning tasks.
Nat. Commun.
(2025)
\end{botherref}
\endbibitem

%%% 12
\bibitem[\protect\citeauthoryear{Rafayelyan et~al.}{2020}]{rafayelyan2020large}
\begin{barticle}
\bauthor{\bsnm{Rafayelyan}, \binits{M.}},
\bauthor{\bsnm{Dong}, \binits{J.}},
\bauthor{\bsnm{Tan}, \binits{Y.}},
\bauthor{\bsnm{Krzakala}, \binits{F.}},
\bauthor{\bsnm{Gigan}, \binits{S.}}:
\batitle{Large-scale optical reservoir computing for spatiotemporal chaotic systems prediction}.
\bjtitle{Phys. Rev. X}
\bvolume{10},
\bfpage{041037}
(\byear{2020})
\end{barticle}
\endbibitem

%%% 13
\bibitem[\protect\citeauthoryear{Sunada and Uchida}{2019}]{sunada2019photonic}
\begin{barticle}
\bauthor{\bsnm{Sunada}, \binits{S.}},
\bauthor{\bsnm{Uchida}, \binits{A.}}:
\batitle{Photonic reservoir computing based on nonlinear wave dynamics at microscale}.
\bjtitle{Sci. Rep.}
\bvolume{9},
\bfpage{19078}
(\byear{2019})
\end{barticle}
\endbibitem

%%% 14
\bibitem[\protect\citeauthoryear{Larger et~al.}{2017}]{larger2017high}
\begin{barticle}
\bauthor{\bsnm{Larger}, \binits{L.}},
\bauthor{\bsnm{Bayl{\'o}n-Fuentes}, \binits{A.}},
\bauthor{\bsnm{Martinenghi}, \binits{R.}},
\bauthor{\bsnm{Udaltsov}, \binits{V.S.}},
\bauthor{\bsnm{Chembo}, \binits{Y.K.}},
\bauthor{\bsnm{Jacquot}, \binits{M.}}:
\batitle{High-speed photonic reservoir computing using a time-delay-based architecture: Million words per second classification}.
\bjtitle{Phys. Rev. X}
\bvolume{7},
\bfpage{011015}
(\byear{2017})
\end{barticle}
\endbibitem

%%% 15
\bibitem[\protect\citeauthoryear{Brunner et~al.}{2013}]{brunner2013parallel}
\begin{barticle}
\bauthor{\bsnm{Brunner}, \binits{D.}},
\bauthor{\bsnm{Soriano}, \binits{M.C.}},
\bauthor{\bsnm{Mirasso}, \binits{C.R.}},
\bauthor{\bsnm{Fischer}, \binits{I.}}:
\batitle{Parallel photonic information processing at gigabyte per second data rates using transient states}.
\bjtitle{Nat. Commun.}
\bvolume{4},
\bfpage{1364}
(\byear{2013})
\end{barticle}
\endbibitem

%%% 16
\bibitem[\protect\citeauthoryear{Nishioka et~al.}{2025}]{nishioka2025two}
\begin{barticle}
\bauthor{\bsnm{Nishioka}, \binits{D.}},
\bauthor{\bsnm{Kitano}, \binits{H.}},
\bauthor{\bsnm{Namiki}, \binits{W.}},
\bauthor{\bsnm{Souma}, \binits{S.}},
\bauthor{\bsnm{Terabe}, \binits{K.}},
\bauthor{\bsnm{Tsuchiya}, \binits{T.}}:
\batitle{Two orders of magnitude reduction in computational load achieved by ultrawideband responses of an ion-gating reservoir}.
\bjtitle{ACS Nano}
\bvolume{19},
\bfpage{36896}--\blpage{36914}
(\byear{2025})
\end{barticle}
\endbibitem

%%% 17
\bibitem[\protect\citeauthoryear{Zhong et~al.}{2022}]{zhong2022memristor}
\begin{barticle}
\bauthor{\bsnm{Zhong}, \binits{Y.}},
\bauthor{\bsnm{Tang}, \binits{J.}},
\bauthor{\bsnm{Li}, \binits{X.}},
\bauthor{\bsnm{Liang}, \binits{X.}},
\bauthor{\bsnm{Liu}, \binits{Z.}},
\bauthor{\bsnm{Li}, \binits{Y.}},
\bauthor{\bsnm{Xi}, \binits{Y.}},
\bauthor{\bsnm{Yao}, \binits{P.}},
\bauthor{\bsnm{Hao}, \binits{Z.}},
\bauthor{\bsnm{Gao}, \binits{B.}}, \betal:
\batitle{A memristor-based analogue reservoir computing system for real-time and power-efficient signal processing}.
\bjtitle{Nat. Electron.}
\bvolume{5},
\bfpage{672}--\blpage{681}
(\byear{2022})
\end{barticle}
\endbibitem

%%% 18
\bibitem[\protect\citeauthoryear{Liu et~al.}{2022}]{liu2022multilayer}
\begin{barticle}
\bauthor{\bsnm{Liu}, \binits{K.}},
\bauthor{\bsnm{Dang}, \binits{B.}},
\bauthor{\bsnm{Zhang}, \binits{T.}},
\bauthor{\bsnm{Yang}, \binits{Z.}},
\bauthor{\bsnm{Bao}, \binits{L.}},
\bauthor{\bsnm{Xu}, \binits{L.}},
\bauthor{\bsnm{Cheng}, \binits{C.}},
\bauthor{\bsnm{Huang}, \binits{R.}},
\bauthor{\bsnm{Yang}, \binits{Y.}}:
\batitle{Multilayer reservoir computing based on ferroelectric {$\alpha$}-{In}$_2${Se}$_3$ for hierarchical information processing}.
\bjtitle{Adv. Mater.}
\bvolume{34},
\bfpage{2108826}
(\byear{2022})
\end{barticle}
\endbibitem

%%% 19
\bibitem[\protect\citeauthoryear{Nishioka et~al.}{2022}]{nishioka2022edge}
\begin{barticle}
\bauthor{\bsnm{Nishioka}, \binits{D.}},
\bauthor{\bsnm{Tsuchiya}, \binits{T.}},
\bauthor{\bsnm{Namiki}, \binits{W.}},
\bauthor{\bsnm{Takayanagi}, \binits{M.}},
\bauthor{\bsnm{Imura}, \binits{M.}},
\bauthor{\bsnm{Koide}, \binits{Y.}},
\bauthor{\bsnm{Higuchi}, \binits{T.}},
\bauthor{\bsnm{Terabe}, \binits{K.}}:
\batitle{Edge-of-chaos learning achieved by ion-electron--coupled dynamics in an ion-gating reservoir}.
\bjtitle{Sci. Adv.}
\bvolume{8},
\bfpage{1156}
(\byear{2022})
\end{barticle}
\endbibitem

%%% 20
\bibitem[\protect\citeauthoryear{Zhong et~al.}{2021}]{zhong2021dynamic}
\begin{barticle}
\bauthor{\bsnm{Zhong}, \binits{Y.}},
\bauthor{\bsnm{Tang}, \binits{J.}},
\bauthor{\bsnm{Li}, \binits{X.}},
\bauthor{\bsnm{Gao}, \binits{B.}},
\bauthor{\bsnm{Qian}, \binits{H.}},
\bauthor{\bsnm{Wu}, \binits{H.}}:
\batitle{Dynamic memristor-based reservoir computing for high-efficiency temporal signal processing}.
\bjtitle{Nat. Commun.}
\bvolume{12},
\bfpage{408}
(\byear{2021})
\end{barticle}
\endbibitem

%%% 21
\bibitem[\protect\citeauthoryear{Gartside et~al.}{2022}]{gartside2022reconfigurable}
\begin{barticle}
\bauthor{\bsnm{Gartside}, \binits{J.C.}},
\bauthor{\bsnm{Stenning}, \binits{K.D.}},
\bauthor{\bsnm{Vanstone}, \binits{A.}},
\bauthor{\bsnm{Holder}, \binits{H.H.}},
\bauthor{\bsnm{Arroo}, \binits{D.M.}},
\bauthor{\bsnm{Dion}, \binits{T.}},
\bauthor{\bsnm{Caravelli}, \binits{F.}},
\bauthor{\bsnm{Kurebayashi}, \binits{H.}},
\bauthor{\bsnm{Branford}, \binits{W.R.}}:
\batitle{Reconfigurable training and reservoir computing in an artificial spin-vortex ice via spin-wave fingerprinting}.
\bjtitle{Nat. Nanotechnol.}
\bvolume{17},
\bfpage{460}--\blpage{469}
(\byear{2022})
\end{barticle}
\endbibitem

%%% 22
\bibitem[\protect\citeauthoryear{Lee et~al.}{2024}]{lee2024task}
\begin{barticle}
\bauthor{\bsnm{Lee}, \binits{O.}},
\bauthor{\bsnm{Wei}, \binits{T.}},
\bauthor{\bsnm{Stenning}, \binits{K.D.}},
\bauthor{\bsnm{Gartside}, \binits{J.C.}},
\bauthor{\bsnm{Prestwood}, \binits{D.}},
\bauthor{\bsnm{Seki}, \binits{S.}},
\bauthor{\bsnm{Aqeel}, \binits{A.}},
\bauthor{\bsnm{Karube}, \binits{K.}},
\bauthor{\bsnm{Kanazawa}, \binits{N.}},
\bauthor{\bsnm{Taguchi}, \binits{Y.}}, \betal:
\batitle{Task-adaptive physical reservoir computing}.
\bjtitle{Nat. Mater.}
\bvolume{23},
\bfpage{79}--\blpage{87}
(\byear{2024})
\end{barticle}
\endbibitem

%%% 23
\bibitem[\protect\citeauthoryear{Prychynenko et~al.}{2018}]{prychynenko2018magnetic}
\begin{barticle}
\bauthor{\bsnm{Prychynenko}, \binits{D.}},
\bauthor{\bsnm{Sitte}, \binits{M.}},
\bauthor{\bsnm{Litzius}, \binits{K.}},
\bauthor{\bsnm{Kr{\"u}ger}, \binits{B.}},
\bauthor{\bsnm{Bourianoff}, \binits{G.}},
\bauthor{\bsnm{Kl{\"a}ui}, \binits{M.}},
\bauthor{\bsnm{Sinova}, \binits{J.}},
\bauthor{\bsnm{Everschor-Sitte}, \binits{K.}}:
\batitle{Magnetic skyrmion as a nonlinear resistive element: a potential building block for reservoir computing}.
\bjtitle{Phys. Rev. Appl.}
\bvolume{9},
\bfpage{014034}
(\byear{2018})
\end{barticle}
\endbibitem

%%% 24
\bibitem[\protect\citeauthoryear{Chen et~al.}{2025}]{chen2025spintronic}
\begin{barticle}
\bauthor{\bsnm{Chen}, \binits{J.}},
\bauthor{\bsnm{Song}, \binits{Y.}},
\bauthor{\bsnm{Hirose}, \binits{A.}}:
\batitle{Spintronic reservoir computing with interpretable nonlinearity}.
\bjtitle{Phys. Rev. Res.}
\bvolume{7},
\bfpage{013310}
(\byear{2025})
\end{barticle}
\endbibitem

%%% 25
\bibitem[\protect\citeauthoryear{Tsunegi et~al.}{2019}]{tsunegi2019physical}
\begin{barticle}
\bauthor{\bsnm{Tsunegi}, \binits{S.}},
\bauthor{\bsnm{Taniguchi}, \binits{T.}},
\bauthor{\bsnm{Nakajima}, \binits{K.}},
\bauthor{\bsnm{Miwa}, \binits{S.}},
\bauthor{\bsnm{Yakushiji}, \binits{K.}},
\bauthor{\bsnm{Fukushima}, \binits{A.}},
\bauthor{\bsnm{Yuasa}, \binits{S.}},
\bauthor{\bsnm{Kubota}, \binits{H.}}:
\batitle{Physical reservoir computing based on spin torque oscillator with forced synchronization}.
\bjtitle{Appl. Phys. Lett.}
\bvolume{114},
\bfpage{163501}
(\byear{2019})
\end{barticle}
\endbibitem

%%% 26
\bibitem[\protect\citeauthoryear{Kurebayashi et~al.}{2026}]{kurebayashi2026metrics}
\begin{barticle}
\bauthor{\bsnm{Kurebayashi}, \binits{H.}},
\bauthor{\bsnm{Finocchio}, \binits{G.}},
\bauthor{\bsnm{Everschor-Sitte}, \binits{K.}},
\bauthor{\bsnm{Gartside}, \binits{J.C.}},
\bauthor{\bsnm{Taniguchi}, \binits{T.}},
\bauthor{\bsnm{Litvinenko}, \binits{A.}},
\bauthor{\bsnm{Kumar}, \binits{A.}},
\bauthor{\bsnm{{\AA}kerman}, \binits{J.}},
\bauthor{\bsnm{Vasilaki}, \binits{E.}},
\bauthor{\bsnm{Sel{\c{c}}uk}, \binits{K.}}, \betal:
\batitle{Metrics for spin-based computing}.
\bjtitle{Nat. Rev. Phys.}
\bvolume{8},
\bfpage{1}--\blpage{18}
(\byear{2026})
\end{barticle}
\endbibitem

%%% 27
\bibitem[\protect\citeauthoryear{Lifshitz and Pitaevskii}{1980}]{landau1980statistical}
\begin{bbook}
\bauthor{\bsnm{Lifshitz}, \binits{E.M.}},
\bauthor{\bsnm{Pitaevskii}, \binits{L.P.}}:
\bbtitle{Statistical Physics: Theory of the Condensed State}.
\bsertitle{Course of Theoretical Physics},
vol. \bseriesno{9}.
\bpublisher{Pergamon},
\blocation{Oxford}
(\byear{1980})
\end{bbook}
\endbibitem

%%% 28
\bibitem[\protect\citeauthoryear{Strogatz}{2015}]{strogatz2024nonlinear}
\begin{bbook}
\bauthor{\bsnm{Strogatz}, \binits{S.H.}}:
\bbtitle{Nonlinear Dynamics and Chaos: With Applications to Physics, Biology, Chemistry, and Engineering},
\bedition{2nd} edn.
\bpublisher{CRC Press},
\blocation{Boca Raton, FL}
(\byear{2015})
\end{bbook}
\endbibitem

%%% 29
\bibitem[\protect\citeauthoryear{K{\"o}rber et~al.}{2023}]{korber2023pattern}
\begin{barticle}
\bauthor{\bsnm{K{\"o}rber}, \binits{L.}},
\bauthor{\bsnm{Heins}, \binits{C.}},
\bauthor{\bsnm{Hula}, \binits{T.}},
\bauthor{\bsnm{Kim}, \binits{J.-V.}},
\bauthor{\bsnm{Thlang}, \binits{S.}},
\bauthor{\bsnm{Schultheiss}, \binits{H.}},
\bauthor{\bsnm{Fassbender}, \binits{J.}},
\bauthor{\bsnm{Schultheiss}, \binits{K.}}:
\batitle{Pattern recognition in reciprocal space with a magnon-scattering reservoir}.
\bjtitle{Nat. Commun.}
\bvolume{14},
\bfpage{3954}
(\byear{2023})
\end{barticle}
\endbibitem

%%% 30
\bibitem[\protect\citeauthoryear{Butschek et~al.}{2022}]{butschek2022photonic}
\begin{barticle}
\bauthor{\bsnm{Butschek}, \binits{L.}},
\bauthor{\bsnm{Akrout}, \binits{A.}},
\bauthor{\bsnm{Dimitriadou}, \binits{E.}},
\bauthor{\bsnm{Lupo}, \binits{A.}},
\bauthor{\bsnm{Haelterman}, \binits{M.}},
\bauthor{\bsnm{Massar}, \binits{S.}}:
\batitle{Photonic reservoir computer based on frequency multiplexing}.
\bjtitle{Opt. Lett.}
\bvolume{47},
\bfpage{782}--\blpage{785}
(\byear{2022})
\end{barticle}
\endbibitem

%%% 31
\bibitem[\protect\citeauthoryear{Lupo et~al.}{2023}]{lupo2023deep}
\begin{barticle}
\bauthor{\bsnm{Lupo}, \binits{A.}},
\bauthor{\bsnm{Picco}, \binits{E.}},
\bauthor{\bsnm{Zajnulina}, \binits{M.}},
\bauthor{\bsnm{Massar}, \binits{S.}}:
\batitle{Deep photonic reservoir computer based on frequency multiplexing with fully analog connection between layers}.
\bjtitle{Optica}
\bvolume{10},
\bfpage{1478}--\blpage{1485}
(\byear{2023})
\end{barticle}
\endbibitem

%%% 32
\bibitem[\protect\citeauthoryear{Namiki et~al.}{2025}]{namiki2025iono}
\begin{barticle}
\bauthor{\bsnm{Namiki}, \binits{W.}},
\bauthor{\bsnm{Nishioka}, \binits{D.}},
\bauthor{\bsnm{Nomura}, \binits{Y.}},
\bauthor{\bsnm{Tsuchiya}, \binits{T.}},
\bauthor{\bsnm{Yamamoto}, \binits{K.}},
\bauthor{\bsnm{Terabe}, \binits{K.}}:
\batitle{Iono--magnonic reservoir computing with chaotic spin wave interference manipulated by ion-gating}.
\bjtitle{Adv. Sci.}
\bvolume{12},
\bfpage{2411777}
(\byear{2025})
\end{barticle}
\endbibitem

%%% 33
\bibitem[\protect\citeauthoryear{Watt and Kostylev}{2020}]{watt2020reservoir}
\begin{barticle}
\bauthor{\bsnm{Watt}, \binits{S.}},
\bauthor{\bsnm{Kostylev}, \binits{M.}}:
\batitle{Reservoir computing using a spin-wave delay-line active-ring resonator based on yttrium-iron-garnet film}.
\bjtitle{Phys. Rev. Appl.}
\bvolume{13},
\bfpage{034057}
(\byear{2020})
\end{barticle}
\endbibitem

%%% 34
\bibitem[\protect\citeauthoryear{Gurevich and Melkov}{1996}]{gurevich2020magnetization}
\begin{bbook}
\bauthor{\bsnm{Gurevich}, \binits{A.G.}},
\bauthor{\bsnm{Melkov}, \binits{G.A.}}:
\bbtitle{Magnetization Oscillations and Waves}.
\bpublisher{CRC Press},
\blocation{Boca Raton, FL}
(\byear{1996})
\end{bbook}
\endbibitem

%%% 35
\bibitem[\protect\citeauthoryear{Bauer et~al.}{2015}]{bauer2015nonlinear}
\begin{barticle}
\bauthor{\bsnm{Bauer}, \binits{H.G.}},
\bauthor{\bsnm{Majchrak}, \binits{P.}},
\bauthor{\bsnm{Kachel}, \binits{T.}},
\bauthor{\bsnm{Back}, \binits{C.H.}},
\bauthor{\bsnm{Woltersdorf}, \binits{G.}}:
\batitle{Nonlinear spin-wave excitations at low magnetic bias fields}.
\bjtitle{Nat. Commun.}
\bvolume{6},
\bfpage{8274}
(\byear{2015})
\end{barticle}
\endbibitem

%%% 36
\bibitem[\protect\citeauthoryear{Krivosik and Patton}{2010}]{krivosik2010hamiltonian}
\begin{barticle}
\bauthor{\bsnm{Krivosik}, \binits{P.}},
\bauthor{\bsnm{Patton}, \binits{C.E.}}:
\batitle{Hamiltonian formulation of nonlinear spin-wave dynamics: theory and applications}.
\bjtitle{Phys. Rev. B}
\bvolume{82},
\bfpage{184428}
(\byear{2010})
\end{barticle}
\endbibitem

%%% 37
\bibitem[\protect\citeauthoryear{Stancil and Prabhakar}{2009}]{stancil2009spin}
\begin{bbook}
\bauthor{\bsnm{Stancil}, \binits{D.D.}},
\bauthor{\bsnm{Prabhakar}, \binits{A.}}:
\bbtitle{Spin Waves: Theory and Applications}.
\bpublisher{Springer},
\blocation{New York, NY}
(\byear{2009})
\end{bbook}
\endbibitem

%%% 38
\bibitem[\protect\citeauthoryear{Chen et~al.}{2026}]{chen2026nonlinear}
\begin{barticle}
\bauthor{\bsnm{Chen}, \binits{J.}},
\bauthor{\bsnm{Song}, \binits{Y.}},
\bauthor{\bsnm{Hirose}, \binits{A.}}:
\batitle{Analysis of nonlinear harmonics in exchange-dominated spin waves}.
\bjtitle{Phys. Rev. B}
\bvolume{113},
\bfpage{064421}
(\byear{2026})
\end{barticle}
\endbibitem

%%% 39
\bibitem[\protect\citeauthoryear{Chumak et~al.}{2022}]{chumak2022advances}
\begin{barticle}
\bauthor{\bsnm{Chumak}, \binits{A.V.}},
\bauthor{\bsnm{Kabos}, \binits{P.}},
\bauthor{\bsnm{Wu}, \binits{M.}},
\bauthor{\bsnm{Abert}, \binits{C.}},
\bauthor{\bsnm{Adelmann}, \binits{C.}},
\bauthor{\bsnm{Adeyeye}, \binits{A.O.}},
\bauthor{\bsnm{{\AA}kerman}, \binits{J.}},
\bauthor{\bsnm{Aliev}, \binits{F.G.}},
\bauthor{\bsnm{Anane}, \binits{A.}},
\bauthor{\bsnm{Awad}, \binits{A.}}, \betal:
\batitle{Advances in magnetics roadmap on spin-wave computing}.
\bjtitle{IEEE Trans. Magn.}
\bvolume{58},
\bfpage{1}--\blpage{72}
(\byear{2022})
\end{barticle}
\endbibitem

%%% 40
\bibitem[\protect\citeauthoryear{Mahmoud et~al.}{2020}]{mahmoud2020introduction}
\begin{barticle}
\bauthor{\bsnm{Mahmoud}, \binits{A.}},
\bauthor{\bsnm{Ciubotaru}, \binits{F.}},
\bauthor{\bsnm{Vanderveken}, \binits{F.}},
\bauthor{\bsnm{Chumak}, \binits{A.V.}},
\bauthor{\bsnm{Hamdioui}, \binits{S.}},
\bauthor{\bsnm{Adelmann}, \binits{C.}},
\bauthor{\bsnm{Cotofana}, \binits{S.}}:
\batitle{Introduction to spin wave computing}.
\bjtitle{J. Appl. Phys.}
\bvolume{128},
\bfpage{161101}
(\byear{2020})
\end{barticle}
\endbibitem

%%% 41
\bibitem[\protect\citeauthoryear{Nakane et~al.}{2018}]{nakane2018reservoir}
\begin{barticle}
\bauthor{\bsnm{Nakane}, \binits{R.}},
\bauthor{\bsnm{Tanaka}, \binits{G.}},
\bauthor{\bsnm{Hirose}, \binits{A.}}:
\batitle{Reservoir computing with spin waves excited in a garnet film}.
\bjtitle{IEEE Access}
\bvolume{6},
\bfpage{4462}--\blpage{4469}
(\byear{2018})
\end{barticle}
\endbibitem

%%% 42
\bibitem[\protect\citeauthoryear{Namiki et~al.}{2023}]{namiki2023experimental}
\begin{barticle}
\bauthor{\bsnm{Namiki}, \binits{W.}},
\bauthor{\bsnm{Nishioka}, \binits{D.}},
\bauthor{\bsnm{Yamaguchi}, \binits{Y.}},
\bauthor{\bsnm{Tsuchiya}, \binits{T.}},
\bauthor{\bsnm{Higuchi}, \binits{T.}},
\bauthor{\bsnm{Terabe}, \binits{K.}}:
\batitle{Experimental demonstration of high-performance physical reservoir computing with nonlinear interfered spin wave multidetection}.
\bjtitle{Adv. Intell. Syst.}
\bvolume{5},
\bfpage{2300228}
(\byear{2023})
\end{barticle}
\endbibitem

%%% 43
\bibitem[\protect\citeauthoryear{Chen et~al.}{2025}]{chen2025analytic}
\begin{barticle}
\bauthor{\bsnm{Chen}, \binits{J.}},
\bauthor{\bsnm{Song}, \binits{Y.}},
\bauthor{\bsnm{Hirose}, \binits{A.}}:
\batitle{Analytic-signal-based input-output modeling inspires passband signal learning for spin-wave reservoir computing}.
\bjtitle{Phys. Rev. Appl.}
\bvolume{23},
\bfpage{034045}
(\byear{2025})
\end{barticle}
\endbibitem

%%% 44
\bibitem[\protect\citeauthoryear{Tanaka et~al.}{2022}]{tanaka2022self}
\begin{barticle}
\bauthor{\bsnm{Tanaka}, \binits{K.}},
\bauthor{\bsnm{Tokudome}, \binits{Y.}},
\bauthor{\bsnm{Minami}, \binits{Y.}},
\bauthor{\bsnm{Honda}, \binits{S.}},
\bauthor{\bsnm{Nakajima}, \binits{T.}},
\bauthor{\bsnm{Takei}, \binits{K.}},
\bauthor{\bsnm{Nakajima}, \binits{K.}}:
\batitle{Self-organization of remote reservoirs: transferring computation to spatially distant locations}.
\bjtitle{Adv. Intell. Syst.}
\bvolume{4},
\bfpage{2100166}
(\byear{2022})
\end{barticle}
\endbibitem

%%% 45
\bibitem[\protect\citeauthoryear{Lee et~al.}{2023}]{lee2023physical}
\begin{barticle}
\bauthor{\bsnm{Lee}, \binits{G.}},
\bauthor{\bsnm{Kang}, \binits{C.}},
\bauthor{\bsnm{Kim}, \binits{S.}},
\bauthor{\bsnm{Park}, \binits{Y.}},
\bauthor{\bsnm{Shin}, \binits{E.J.}},
\bauthor{\bsnm{Cho}, \binits{B.J.}}:
\batitle{Physical reservoir based on a leaky-fefet using the temporal memory effect}.
\bjtitle{IEEE Electron Device Lett.}
\bvolume{45},
\bfpage{108}--\blpage{111}
(\byear{2023})
\end{barticle}
\endbibitem

%%% 46
\bibitem[\protect\citeauthoryear{Toprasertpong et~al.}{2022}]{toprasertpong2022reservoir}
\begin{barticle}
\bauthor{\bsnm{Toprasertpong}, \binits{K.}},
\bauthor{\bsnm{Nako}, \binits{E.}},
\bauthor{\bsnm{Wang}, \binits{Z.}},
\bauthor{\bsnm{Nakane}, \binits{R.}},
\bauthor{\bsnm{Takenaka}, \binits{M.}},
\bauthor{\bsnm{Takagi}, \binits{S.}}:
\batitle{Reservoir computing on a silicon platform with a ferroelectric field-effect transistor}.
\bjtitle{Commun. Eng.}
\bvolume{1},
\bfpage{21}
(\byear{2022})
\end{barticle}
\endbibitem

%%% 47
\bibitem[\protect\citeauthoryear{Rajib et~al.}{2025}]{rajib2025magneto}
\begin{barticle}
\bauthor{\bsnm{Rajib}, \binits{M.M.}},
\bauthor{\bsnm{Bhattacharya}, \binits{D.}},
\bauthor{\bsnm{Jensen}, \binits{C.J.}},
\bauthor{\bsnm{Chen}, \binits{G.}},
\bauthor{\bsnm{Chowdhury}, \binits{F.F.}},
\bauthor{\bsnm{Sarker}, \binits{S.}},
\bauthor{\bsnm{Liu}, \binits{K.}},
\bauthor{\bsnm{Atulasimha}, \binits{J.}}:
\batitle{Magneto-ionic physical reservoir computing in perpendicularly magnetized heterostructures}.
\bjtitle{Nano Lett.}
\bvolume{25},
\bfpage{15369}--\blpage{15376}
(\byear{2025})
\end{barticle}
\endbibitem

%%% 48
\bibitem[\protect\citeauthoryear{Heins et~al.}{2025}]{heins2025benchmarking}
\begin{barticle}
\bauthor{\bsnm{Heins}, \binits{C.}},
\bauthor{\bsnm{Kim}, \binits{J.-V.}},
\bauthor{\bsnm{K{\"o}rber}, \binits{L.}},
\bauthor{\bsnm{Fassbender}, \binits{J.}},
\bauthor{\bsnm{Schultheiss}, \binits{H.}},
\bauthor{\bsnm{Schultheiss}, \binits{K.}}:
\batitle{Benchmarking a magnon-scattering reservoir with modal and temporal multiplexing}.
\bjtitle{Phys. Rev. Appl.}
\bvolume{23},
\bfpage{054087}
(\byear{2025})
\end{barticle}
\endbibitem

%%% 49
\bibitem[\protect\citeauthoryear{Nagase et~al.}{2024}]{nagase2024spin}
\begin{barticle}
\bauthor{\bsnm{Nagase}, \binits{S.}},
\bauthor{\bsnm{Nezu}, \binits{S.}},
\bauthor{\bsnm{Sekiguchi}, \binits{K.}}:
\batitle{Spin-wave reservoir chips with short-term memory for high-speed estimation of external magnetic fields}.
\bjtitle{Phys. Rev. Appl.}
\bvolume{22},
\bfpage{024072}
(\byear{2024})
\end{barticle}
\endbibitem

%%% 50
\bibitem[\protect\citeauthoryear{Namiki et~al.}{2024}]{namiki2024opto}
\begin{barticle}
\bauthor{\bsnm{Namiki}, \binits{W.}},
\bauthor{\bsnm{Yamaguchi}, \binits{Y.}},
\bauthor{\bsnm{Nishioka}, \binits{D.}},
\bauthor{\bsnm{Tsuchiya}, \binits{T.}},
\bauthor{\bsnm{Terabe}, \binits{K.}}:
\batitle{Opto-magnonic reservoir computing coupling nonlinear interfered spin wave and visible light switching}.
\bjtitle{Mater. Today Phys.}
\bvolume{45},
\bfpage{101465}
(\byear{2024})
\end{barticle}
\endbibitem

%%% 51
\bibitem[\protect\citeauthoryear{Akashi et~al.}{2020}]{akashi2020input}
\begin{barticle}
\bauthor{\bsnm{Akashi}, \binits{N.}},
\bauthor{\bsnm{Yamaguchi}, \binits{T.}},
\bauthor{\bsnm{Tsunegi}, \binits{S.}},
\bauthor{\bsnm{Taniguchi}, \binits{T.}},
\bauthor{\bsnm{Nishida}, \binits{M.}},
\bauthor{\bsnm{Sakurai}, \binits{R.}},
\bauthor{\bsnm{Wakao}, \binits{Y.}},
\bauthor{\bsnm{Nakajima}, \binits{K.}}:
\batitle{Input-driven bifurcations and information processing capacity in spintronics reservoirs}.
\bjtitle{Phys. Rev. Res.}
\bvolume{2},
\bfpage{043303}
(\byear{2020})
\end{barticle}
\endbibitem

%%% 52
\bibitem[\protect\citeauthoryear{Kan et~al.}{2021}]{kan2021simple}
\begin{barticle}
\bauthor{\bsnm{Kan}, \binits{S.}},
\bauthor{\bsnm{Nakajima}, \binits{K.}},
\bauthor{\bsnm{Takeshima}, \binits{Y.}},
\bauthor{\bsnm{Asai}, \binits{T.}},
\bauthor{\bsnm{Kuwahara}, \binits{Y.}},
\bauthor{\bsnm{Akai-Kasaya}, \binits{M.}}:
\batitle{Simple reservoir computing capitalizing on the nonlinear response of materials: theory and physical implementations}.
\bjtitle{Phys. Rev. Appl.}
\bvolume{15},
\bfpage{024030}
(\byear{2021})
\end{barticle}
\endbibitem

%%% 53
\bibitem[\protect\citeauthoryear{Akai-Kasaya et~al.}{2022}]{akai2022performance}
\begin{barticle}
\bauthor{\bsnm{Akai-Kasaya}, \binits{M.}},
\bauthor{\bsnm{Takeshima}, \binits{Y.}},
\bauthor{\bsnm{Kan}, \binits{S.}},
\bauthor{\bsnm{Nakajima}, \binits{K.}},
\bauthor{\bsnm{Oya}, \binits{T.}},
\bauthor{\bsnm{Asai}, \binits{T.}}:
\batitle{Performance of reservoir computing in a random network of single-walled carbon nanotubes complexed with polyoxometalate}.
\bjtitle{Neuromorph. Comput. Eng.}
\bvolume{2},
\bfpage{014003}
(\byear{2022})
\end{barticle}
\endbibitem

%%% 54
\bibitem[\protect\citeauthoryear{Yamada et~al.}{2024}]{yamada2024physical}
\begin{barticle}
\bauthor{\bsnm{Yamada}, \binits{R.}},
\bauthor{\bsnm{Nakagawa}, \binits{M.}},
\bauthor{\bsnm{Hirooka}, \binits{S.}},
\bauthor{\bsnm{Tada}, \binits{H.}}:
\batitle{Physical reservoir computing with visible-light signals using dye-sensitized solar cells}.
\bjtitle{Appl. Phys. Express}
\bvolume{17},
\bfpage{097001}
(\byear{2024})
\end{barticle}
\endbibitem

%%% 55
\bibitem[\protect\citeauthoryear{Kan et~al.}{2022}]{kan2022physical}
\begin{barticle}
\bauthor{\bsnm{Kan}, \binits{S.}},
\bauthor{\bsnm{Nakajima}, \binits{K.}},
\bauthor{\bsnm{Asai}, \binits{T.}},
\bauthor{\bsnm{Akai-Kasaya}, \binits{M.}}:
\batitle{Physical implementation of reservoir computing through electrochemical reaction}.
\bjtitle{Adv. Sci.}
\bvolume{9},
\bfpage{2104076}
(\byear{2022})
\end{barticle}
\endbibitem

%%% 56
\bibitem[\protect\citeauthoryear{{Linguistic Data Consortium}}{1993}]{LDC93S9}
\begin{botherref}
\oauthor{\bsnm{{Linguistic Data Consortium}}}:
TI-46 Word Speech Corpus.
Catalog No. LDC93S9, Philadelphia
(1993)
\end{botherref}
\endbibitem

%%% 57
\bibitem[\protect\citeauthoryear{Zolfagharinejad et~al.}{2025}]{zolfagharinejad2025analogue}
\begin{barticle}
\bauthor{\bsnm{Zolfagharinejad}, \binits{M.}},
\bauthor{\bsnm{B{\"u}chel}, \binits{J.}},
\bauthor{\bsnm{Cassola}, \binits{L.}},
\bauthor{\bsnm{Kinge}, \binits{S.}},
\bauthor{\bsnm{Syed}, \binits{G.S.}},
\bauthor{\bsnm{Sebastian}, \binits{A.}},
\bauthor{\bsnm{Wiel}, \binits{W.G.}}:
\batitle{Analogue speech recognition based on physical computing}.
\bjtitle{Nature}
\bvolume{645},
\bfpage{886}--\blpage{892}
(\byear{2025})
\end{barticle}
\endbibitem

%%% 58
\bibitem[\protect\citeauthoryear{Abreu~Araujo et~al.}{2020}]{abreu2020role}
\begin{barticle}
\bauthor{\bsnm{Abreu~Araujo}, \binits{F.}},
\bauthor{\bsnm{Riou}, \binits{M.}},
\bauthor{\bsnm{Torrejon}, \binits{J.}},
\bauthor{\bsnm{Tsunegi}, \binits{S.}},
\bauthor{\bsnm{Querlioz}, \binits{D.}},
\bauthor{\bsnm{Yakushiji}, \binits{K.}},
\bauthor{\bsnm{Fukushima}, \binits{A.}},
\bauthor{\bsnm{Kubota}, \binits{H.}},
\bauthor{\bsnm{Yuasa}, \binits{S.}},
\bauthor{\bsnm{Stiles}, \binits{M.D.}}, \betal:
\batitle{Role of non-linear data processing on speech recognition task in the framework of reservoir computing}.
\bjtitle{Sci. Rep.}
\bvolume{10},
\bfpage{328}
(\byear{2020})
\end{barticle}
\endbibitem

%%% 59
\bibitem[\protect\citeauthoryear{Dang et~al.}{2026}]{dang2026spiking}
\begin{botherref}
\oauthor{\bsnm{Dang}, \binits{B.}},
\oauthor{\bsnm{Zhang}, \binits{T.}},
\oauthor{\bsnm{Meng}, \binits{F.}},
\oauthor{\bsnm{Liu}, \binits{K.}},
\oauthor{\bsnm{Yu}, \binits{L.}},
\oauthor{\bsnm{Zhang}, \binits{Q.}},
\oauthor{\bsnm{Wu}, \binits{S.}},
\oauthor{\bsnm{Gu}, \binits{L.}},
\oauthor{\bsnm{Huang}, \binits{R.}},
\oauthor{\bsnm{Yang}, \binits{Y.}}:
Spiking neural networks with fatigue spike-timing-dependent plasticity learning using hybrid memristor arrays.
Nat. Electron.
(2026)
\end{botherref}
\endbibitem

%%% 60
\bibitem[\protect\citeauthoryear{Cai et~al.}{2023}]{cai2023brain}
\begin{barticle}
\bauthor{\bsnm{Cai}, \binits{H.}},
\bauthor{\bsnm{Ao}, \binits{Z.}},
\bauthor{\bsnm{Tian}, \binits{C.}},
\bauthor{\bsnm{Wu}, \binits{Z.}},
\bauthor{\bsnm{Liu}, \binits{H.}},
\bauthor{\bsnm{Tchieu}, \binits{J.}},
\bauthor{\bsnm{Gu}, \binits{M.}},
\bauthor{\bsnm{Mackie}, \binits{K.}},
\bauthor{\bsnm{Guo}, \binits{F.}}:
\batitle{Brain organoid reservoir computing for artificial intelligence}.
\bjtitle{Nat. Electron.}
\bvolume{6},
\bfpage{1032}--\blpage{1039}
(\byear{2023})
\end{barticle}
\endbibitem

%%% 61
\bibitem[\protect\citeauthoryear{Dion et~al.}{2018}]{dion2018reservoir}
\begin{barticle}
\bauthor{\bsnm{Dion}, \binits{G.}},
\bauthor{\bsnm{Mejaouri}, \binits{S.}},
\bauthor{\bsnm{Sylvestre}, \binits{J.}}:
\batitle{Reservoir computing with a single delay-coupled nonlinear mechanical oscillator}.
\bjtitle{J. Appl. Phys.}
\bvolume{124},
\bfpage{152132}
(\byear{2018})
\end{barticle}
\endbibitem

%%% 62
\bibitem[\protect\citeauthoryear{Govia et~al.}{2021}]{govia2021quantum}
\begin{barticle}
\bauthor{\bsnm{Govia}, \binits{L.C.G.}},
\bauthor{\bsnm{Ribeill}, \binits{G.J.}},
\bauthor{\bsnm{Rowlands}, \binits{G.E.}},
\bauthor{\bsnm{Krovi}, \binits{H.K.}},
\bauthor{\bsnm{Ohki}, \binits{T.A.}}:
\batitle{Quantum reservoir computing with a single nonlinear oscillator}.
\bjtitle{Phys. Rev. Res.}
\bvolume{3},
\bfpage{013077}
(\byear{2021})
\end{barticle}
\endbibitem

%%% 63
\bibitem[\protect\citeauthoryear{del Hougne and Lerosey}{2018}]{del2018leveraging}
\begin{barticle}
\bauthor{\bsnm{Hougne}, \binits{P.}},
\bauthor{\bsnm{Lerosey}, \binits{G.}}:
\batitle{Leveraging chaos for wave-based analog computation: demonstration with indoor wireless communication signals}.
\bjtitle{Phys. Rev. X}
\bvolume{8},
\bfpage{041037}
(\byear{2018})
\end{barticle}
\endbibitem

%%% 64
\bibitem[\protect\citeauthoryear{Im et~al.}{2020}]{im2020memristive}
\begin{barticle}
\bauthor{\bsnm{Im}, \binits{I.H.}},
\bauthor{\bsnm{Kim}, \binits{S.J.}},
\bauthor{\bsnm{Jang}, \binits{H.W.}}:
\batitle{Memristive devices for new computing paradigms}.
\bjtitle{Adv. Intell. Syst.}
\bvolume{2},
\bfpage{2000105}
(\byear{2020})
\end{barticle}
\endbibitem

\end{thebibliography}
